\documentclass[aip,jmp,amsmath,amssymb]{revtex4-1}
\draft

\makeatletter \@addtoreset{equation}{section} \makeatother

\def\dfrac{\displaystyle\frac}
\def\sumd{\displaystyle\sum}

\def\intd{\displaystyle\int}
\def\prodd{\displaystyle\prod}

\newtheorem{theorem}{Theorem}

\newtheorem{proposition}[theorem]{Proposition}

\newtheorem{lemma}[theorem]{Lemma}
\newtheorem{remark}[theorem]{Remark}

\begin{document}

\title{Asymptotic behavior of the Verblunsky coefficients for the OPUC with a varying weight.}
\author{M.Poplavskyi}
\email{poplavskiymihail@rambler.ru}
\affiliation{Mathematical Division, B.Verkin Institute for Low Temperature Physics National Academy of Sciences of Ukraine, 47 Lenin Ave., Kharkiv 61103, Ukraine}

\date{\today}

\begin{abstract}
We present an asymptotic analysis of the Verblunsky coefficients for the polynomials orthogonal on the
unit circle with the varying weight $e^{-nV(\cos x)}$, assuming that the potential $V$
has four bounded derivatives on $[-1,1]$ and the equilibrium measure has a one interval support.
We obtain the asymptotics as a solution of the system of "string" equations.
\end{abstract}

\keywords{Polynomials orthogonal on the unit circle, unitary matrix models, Verblunsky coefficients, CMV matrices.}


\maketitle

\section{Introduction}\label{s:Intr}
In this paper we consider a systems of polynomials orthogonal on the unit circle (OPUC) with a varying weight. The system can be obtained from $\left\{e^{i k \lambda}\right\}_{k=0}^{\infty}$ if we use the Gram-Shmidt
procedure in $L^{\left(n\right)}:=L_2\left(
\left[-\pi,\pi\right], e^{-nV\left(\cos \lambda\right)} \right)$ with the inner product
\begin{equation*}
\left\langle f,g\right\rangle_n=\intd_{-\pi}^{\pi} f\left(x\right)
\overline{g\left(x\right)} e^{-nV\left(\cos x\right)}dx.
\end{equation*}
Then for any $n$ we get the system
of trigonometric polynomials
\begin{equation*}\label{d:P_def}
P_k^{\left(n\right)}\left(\lambda\right) =
\sumd_{l=0}^{k}
c_{k,l}^{\left(n\right)} e^{il\lambda},
\end{equation*}
which are orthonormal in $L^{\left(n\right)}$.
One can see from the Szeg$\ddot{o}$'s condition that the system
$\left\{P_k^{\left(n\right)}\left(\lambda\right)\right\}_{k=0}^{\infty}$ is not complete in $L^{\left(n\right)}$.
To construct the complete system one should include also polynomials with respect to $e^{-i\lambda}$.
Hence following [\onlinecite{CMV:03}] we consider reversed polynomials
\begin{equation*}\label{d:Q_def}
Q_k^{\left(n\right)}\left(\lambda\right)=e^{ik\lambda}
P_k^{\left(n\right)}\left(-\lambda\right)= \sumd_{l=0}^{k}
c_{k,l}^{\left(n\right)} e^{i\left(k-l\right)\lambda},
\end{equation*}
and define the set of functions $\left\{\chi_k^{\left(n\right)}\left(\lambda\right)\right\}_{k=0}^{\infty}$
\begin{eqnarray}
\chi_{2k}^{\left(n\right)}\left(\lambda\right) =
&e^{-ik\lambda}Q_{2k}^{\left(n\right)}\left(\lambda\right),
\notag
\\
\chi_{2k+1}^{\left(n\right)}\left(\lambda\right)=
&e^{-ik\lambda}P_{2k+1}^{\left(n\right)}\left(\lambda\right).
\label{d:Chi_k}
\end{eqnarray}
It is easy to check (see, e.g [\onlinecite{CMV:03}]) that the system $\left\{\chi_k^{\left(n\right)}
\left(\lambda\right)\right\}_{k=0}^{\infty}$ is an orthonormal basis in $L^{\left(n\right)}$.
 Moreover, it was proven in [\onlinecite{CMV:03}] that the functions $\chi_k^{\left(n\right)}$ satisfy the five term recurrent relations
\begin{eqnarray}
e^{i\lambda}\chi_{2k-1}^{\left(n\right)}\left(\lambda\right)=
&
-\alpha_{2k}^{\left(n\right)}\rho_{2k-1}^{\left(n\right)}
\chi_{2k-2}^{\left(n\right)}\left(\lambda\right)%
-\alpha_{2k}^{\left(n\right)}\alpha_{2k-1}^{\left(n\right)}
\chi_{2k-1}^{\left(n\right)}\left(\lambda\right)%
\notag
\\
& -\alpha_{2k+1}^{\left(n\right)}\rho_{2k}^{\left(n\right)}
\chi_{2k}^{\left(n\right)}\left(\lambda\right)%
+\rho_{2k}^{\left(n\right)}\rho_{2k+1}^{\left(n\right)}
\chi_{2k+1}^{\left(n\right)}\left(\lambda\right),%
\label{e:CMV1}
\\
e^{i\lambda}\chi_{2k}^{\left(n\right)}\left(\lambda\right)=
&
\rho_{2k}^{\left(n\right)}\rho_{2k-1}^{\left(n\right)}
\chi_{2k-2}^{\left(n\right)}\left(\lambda\right)%
+ \alpha_{2k-1}^{\left(n\right)}\rho_{2k}^{\left(n\right)}
\chi_{2k-1}^{\left(n\right)}\left(\lambda\right)%
\notag
\\
& -\alpha_{2k+1}^{\left(n\right)}\alpha_{2k}^{\left(n\right)}
\chi_{2k}^{\left(n\right)}\left(\lambda\right)%
+\alpha_{2k}^{\left(n\right)}\rho_{2k+1}^{\left(n\right)}
\chi_{2k+1}^{\left(n\right)}\left(\lambda\right),%
\label{e:CMV2}
\end{eqnarray}
where $\alpha_k^{\left(n\right)}=c_{k,0}^{\left(n\right)}/c_{k,k}^{\left(n\right)}$ %
and
$\rho_k^{\left(n\right)}=c_{k-1,k-1}^{\left(n\right)}/c_{k,k}^{\left(n\right)}$
are called the Verblunsky coefficients of the system
$\left\{\chi_k^{\left(n\right)}\left(\lambda\right)\right\}_{k=0}^{\infty}$.
They satisfy the relations
\begin{equation*}\label{e:alpha_rho}
\left(\rho_k^{\left(n\right)}\right)^2+\left(\alpha_k^{\left(n\right)}\right)^2=1.
\end{equation*}%
Hence the matrix of the operator $C$ of multiplication by $e^{i\lambda}$ in the basis $\left\{\chi_k^{\left(n\right)}\left(\lambda\right)\right\}_{k=0}^{\infty}$ has the form  (see [\onlinecite{CMV:03}])
\begin{equation}\label{m:CMV}
C^{\left(n\right)}= %
\left(%
\begin{array}{ccccccc}
-\alpha_{1}^{\left(n\right)} & \rho_{1}^{\left(n\right)} & 0 & 0 & 0
& 0 & \ldots
\\
-\rho_{1}^{\left(n\right)}\alpha_{2}^{\left(n\right)} &
-\alpha_{1}^{\left(n\right)}\alpha_{2}^{\left(n\right)} &
-\rho_{2}^{\left(n\right)}\alpha_{3}^{\left(n\right)} &
\rho_{2}^{\left(n\right)}\rho_{3}^{\left(n\right)}  & 0 & 0 &\ldots
\\
\rho_{1}^{\left(n\right)}\rho_{2}^{\left(n\right)} &
\alpha_{1}^{\left(n\right)}\rho_{2}^{\left(n\right)} &
-\alpha_{2}^{\left(n\right)}\alpha_{3}^{\left(n\right)} &
\alpha_{2}^{\left(n\right)}\rho_{3}^{\left(n\right)}  & 0 & 0
&\ldots
\\
0 & 0 & -\rho_{3}^{\left(n\right)}\alpha_{4}^{\left(n\right)} &
-\alpha_{3}^{\left(n\right)}\alpha_{4}^{\left(n\right)} &
-\rho_{4}^{\left(n\right)}\alpha_{5}^{\left(n\right)} &
\rho_{4}^{\left(n\right)}\rho_{5}^{\left(n\right)}  & \ldots
\\
0 & 0 & \rho_{3}^{\left(n\right)}\rho_{4}^{\left(n\right)} &
\alpha_{3}^{\left(n\right)}\rho_{4}^{\left(n\right)} &
-\alpha_{4}^{\left(n\right)}\alpha_{5}^{\left(n\right)} &
\alpha_{4}^{\left(n\right)}\rho_{5}^{\left(n\right)}  & \ldots
\\
0 & 0 & 0 & \ddots & \ddots & \ddots & \ddots
\end{array}
\right).%
\end{equation}
The main goal of the paper is to find the first two terms of the asymptotic expansion of
 $\alpha_{n+k}^{(n)}$ for $k=o\left(n\right),\, n \to \infty$.
The asymptotics of polynomials orthogonal with respect to a varying weight plays an important role
in many problems of modern mathematics, e.g. approximation theory, combinatorics, theory
of random matrices etc. The polynomials orthogonal with respect to a varying weight on the real line are
now well studied, due to the remarkable paper [\onlinecite{De-Co:99}], where it was shown that
for real analytic $V\left(\lambda\right)$ satisfying also some growth conditions as
$\left|\lambda\right|\to \infty$, the asymptotics of polynomials
$P_{n+k}^{(n)}\left(\lambda\right)$ $\left(k=o\left(n\right),\, n\to \infty\right)$
orthogonal on the real line (OPRL) with the weight $e^{-nV\left(\lambda\right)}$, is expressed via $\theta$-functions,
which can be constructed in  terms of the so-called equilibrium measure
$\mu\left(d \lambda\right) = \rho\left(\lambda\right)d \lambda$, that is obtained as a
solution of a classical problem of the potential theory with the potential $V$. These results
were generalized on the case of $V$ possessing only few derivatives in the work [\onlinecite{Mc:08}].
The complete asymptotic expansion of the Jacobi matrix coefficients in the case of real analytic
$V$ and one or two interval equilibrium density $\rho$ was constructed in [\onlinecite{Pa-Sh-Al:01}].
These results were used to prove universality of local bulk and edge regimes of the hermitian matrix models
with the potential $V$ (see also [\onlinecite{Pa-Sh:07}], [\onlinecite{Pa-Sh:03}], [\onlinecite{LL:08}]). The case of OPUC, with
varying weight till now is not understood so well as the case of classical OP. It is believed that similarly to
the real line case the asymptotics is closely connected with the behavior of the corresponding equilibrium measure,
which is the minimizer of the functional
\begin{equation*}\label{e:VP}
\mathcal{E}[m]=\intd_{-\pi}^{\pi} V(\cos \lambda )m(d\lambda )-\intd_{-\pi}^{\pi} \log \left|e^{i \lambda} - e^{i\mu}
\right|m(d\lambda )m(d\mu ),
\end{equation*}
in the class of unit measures on the interval $\left[-\pi,\pi\right]$ (see [\onlinecite{ST:97}]
for existence and properties of the solution). It is well known, in particular, that for smooth $V'$
the equilibrium measure has a density $\rho$ which is uniquely defined by the condition that the function
\begin{equation}\label{d:u}
u\left(\lambda\right)= V\left(\cos\lambda\right)-2\intd_{\sigma} \log \left|e^{i\lambda}-e^{i\mu}\right| \rho\left(\mu\right) d\mu,
\end{equation}
takes it minimum value if $\lambda \in \sigma = \mathrm{supp} \,  \rho$.
From this condition in the case of differentiable $V$ one can obtain the following integral equation for the equilibrium density $\rho$
\begin{equation}\label{e:V_Rho}
\left( V \left( \cos \lambda \right) \right)'=v.p.\intd_{-\pi}^{\pi} \cot
\dfrac{\lambda-\mu}{2} \rho \left(\mu\right) d\mu, \quad
\mbox{for}\, \lambda \, \in \, \sigma.
\end{equation}%
The asymptotics of OPUC with varying weight was constructed in [\onlinecite{Mc:06}] for the case
when the support $\sigma$ of $\rho$ is the whole circle. It was proven that in this case
$P^{(n)}_{n} \left(\lambda\right) \sim C_V e^{i n\lambda}$ and $\alpha_n^{(n)} = \overline{o} \left(1\right)$.
In the present paper we find the first two terms of the asymptotic expansion of $\alpha^{(n)}_{n+k}$
in the case of one-interval
support $\sigma$ of the equilibrium density $\rho$. Our main conditions are

\noindent
\textbf{Condition C1.}
\textit{
The support $\sigma$ of the equilibrium measure
is a single subinterval of the interval $\left[-\pi,\pi\right]$, i.e.
\begin{equation}\label{e:C1}
\sigma = \left[-\theta,\theta\right], \, \text{with} \quad \theta < \pi.
\end{equation}
}

\noindent
\textbf{Condition C2.}
\textit{
The equilibrium density $\rho$ has no zeros in $(-\theta,\theta)$ and
\begin{equation}\label{e:C2}
\rho\left(\lambda\right) \sim C  \left|\lambda \mp \theta\right|^{1/2}, \, \text{for}\; \lambda \rightarrow \pm
\theta,
\end{equation}
and the function $u\left(\lambda\right)$ of \eqref{d:u}
attains its minimum if and only if $\lambda$ belongs to  $\sigma$.
}

\noindent
\textbf{Condition C3.}
\textit{
 $V\left(\cos \lambda\right)$ possesses 4 bounded derivatives on $\sigma_{\varepsilon}=\left[-\theta-\varepsilon,
\theta+\varepsilon\right]$.
}
\begin{remark}
In fact there is one more possibility to have one-interval $\sigma$.
 Another case is a some left symmetric arc of the circle, i.e. $\left[\pi-\theta,\pi+\theta\right]$.
In this case we replace $V\left(\cos x\right)$ by $V\left(\cos\left(\pi-x\right)\right)$. After this replacement we 
obtain the support from the condition C1. The scalar product $\langle e^{ikx},e^{imx}\rangle_n$ will change on factor $\left(-1\right)^{k-m}$.
Therefore, one can see from the Gram-Shmidt algorithm, that coefficients $\alpha_k^{\left(n\right)}$ will change on factor $\left(-1\right)^k$.   
\end{remark}
The following simple representation of $\rho$ plays an important role in our asymptotic analysis
\begin{proposition}\label{p:Rho}
Under conditions C1-C3
\begin{equation}\label{e:Rho_a}
\rho \left(\lambda\right) = \dfrac{1}{4\pi^2}\chi
\left(\lambda\right)P\left(\lambda\right)\mathbf{1_{\sigma}},
\end{equation}
where
\begin{equation}\label{d:Sqrt}
\chi \left(\lambda\right) =
\sqrt{\left|\cos\lambda-\cos\theta\right|}, \quad
P\left(\lambda\right) = \intd_{-\theta}^{\theta} \dfrac{\left(V
\left(\cos \mu\right)\right)' - \left(V \left(\cos
\lambda\right)\right)'}{\sin\left(\mu-\lambda\right)/2}
\dfrac{d\mu}{\chi \left(\mu\right)}.
\end{equation}
\end{proposition}
The proof is very simple but for the reader's convenience we give it in the beginning of Section~\ref{s:Auxillary}.
It is evident that conditions C1-C3 imply that $P\left(\lambda\right)$ does not have roots on interval
$\left[-\theta,\theta\right]$.
\begin{remark}\label{p:Rho_w}
If the support of the equilibrium measure is the whole interval
$\left[-\pi,\pi\right]$, then equation \eqref{e:V_Rho} can be
solved using the Hilbert operator.
\begin{equation}\label{e:Rho_w}
\rho \left(\mu\right)=\dfrac{1}{2\pi}+\dfrac{1}{4\pi^2}
\intd_{-\pi}^{\pi} \cot \dfrac{\lambda-\mu}{2} \left(\left(V
\left(\cos \lambda\right)\right)' - \left(V \left(\cos
\mu\right)\right)' \right)d\lambda.
\end{equation}
\end{remark}
 The main result of the paper is the following theorem:
\begin{theorem}\label{t:Main}
Consider the system of orthogonal polynomials and the Verblunsky
coefficients defined in \eqref{e:CMV1} and \eqref{e:CMV2}. Let
potential $V$ satisfy conditions C1 - C3 above. Then there exists
$\varepsilon_1
> 0$ such that for any $\left|m\right|<\varepsilon_1 n$
\begin{equation}\label{e:MainAs}
\alpha_{n+m}^{\left(n\right)}=\left(-1\right)^{n+m} s \cos
\left(\dfrac{\theta}{2}+x^{\left(n\right)}_{n+m}\right),
\end{equation}
where $ s = 1$ or $ s = -1$ and
\begin{equation*}
x_{n+m} =\dfrac{\pi \sqrt2 }{P\left(\theta\right) \sin\theta/2}
\dfrac{m}{n} + \underline{O} \left(\log^{11} n \left(n^{-4/3}+
\dfrac{m^2}{n^2}\right)\right),
\end{equation*}
with $P$ defined in \eqref{d:Sqrt}.
\end{theorem}
\begin{remark}
The sign of $s$ can not be explicitly found and may be different for different values of $n$.
As we are going to use this result in future for the proving of universality conjecture on the edge of spectrum, we expect that the sign of $s$ will not affect the final result (for the example see [\onlinecite{Pa-Sh:03}]).
\end{remark}
The proof of Theorem~\ref{t:Main} is based on the relation, which is an analog of the
string equation for  OP on the real line
\begin{proposition}\label{p:Str_eq}
Let $\chi_{k}^{\left(n\right)}$ be the orthogonal functions
defined in \eqref{d:Chi_k}. Then for any $k$
\begin{equation}\label{e:Str_eq}
\intd_{-\pi}^{\pi} \left(\sin \lambda \right) V' \left(\cos \lambda\right)
\chi_{k}^{\left(n\right)} \left(\lambda\right)
\overline{\chi_{k-1}^{\left(n\right)}\left(\lambda\right)}
e^{-nV\left(\cos\lambda\right)} d\lambda = i\left(-1\right)^{k-1}
\dfrac{k}{n} \dfrac{\alpha_{k}^{\left(n\right)}}
{\rho_{k}^{\left(n\right)}}.
\end{equation}
\end{proposition}
The result was obtained in [\onlinecite{Hi:96}], but in some different form, therefore we give its proof in Section~\ref{s:Auxillary}.
We use \eqref{e:Str_eq} as a system of nonlinear equations for $\alpha_{k}^{\left(n\right)}$
and solve it by the  perturbation theory method. To construct a zero order solution of the system \eqref{e:Str_eq}, following the method of [\onlinecite{Pa-Sh-Al:01}] and [\onlinecite{Pa-Sh:03}] we use the link of the OPUC with the unitary matrix models, which eigenvalue distribution has the form
\begin{equation}\label{d:PDef}
p_n \left( \lambda_1, \ldots , \lambda_
n\right)=\dfrac{1}{Z_n}\prodd\limits_{1 \leq j < k \leq n} %
\left| e^{i \lambda_j} -  e^{i \lambda_k}\right|^2 %
\exp \left\{ -n \sum\limits_{j=1}^{n} V \left( \cos\lambda_j \right)
\right\}.
\end{equation}
It is known (see [\onlinecite{Me:91}]) that all marginal densities of \eqref{d:PDef} can be expressed using the
orthogonal functions
\begin{equation}\label{d:Psi_k}
\psi_k^{(n)} \left(\lambda\right)= P_k^{(n)} \left(\lambda\right)
e^{-nV\left(\cos \lambda\right)/2}.
\end{equation}
and their reproducing kernel
\begin{equation}\label{d:K_k_n}
K_{k,n} \left(\lambda, \mu\right) = \sumd_{l=0}^{k-1}
\psi_l^{\left(n\right)} \left(\lambda\right)
\overline{\psi_l^{\left(n\right)} \left(\mu\right)}.
\end{equation}
These ensembles first appeared in physics, but later their relationship to the pure mathematical problems was discovered (see, e.g. [\onlinecite{Jo-D:98}]).

The paper is organized as follows. In Section~\ref{s:Basic} we prove the main Theorem~\ref{t:Main}
using some technical results. We will prove these results in Section~\ref{s:Auxillary}.

\section{Proof of  basic results}\label{s:Basic}
The method of the proof is a version of the one used in [\onlinecite{Pa-Sh:03}]. An important part of the proof
is a zero order approximation for the Verblunsky coefficients $\alpha^{\left(n\right)}_{n+k}$.
\begin{theorem}\label{t:NullApp}
Under conditions of Theorem \ref{t:Main} we have
\begin{equation}\label{e:NullApp}
\alpha^{\left(n\right)}_{n+k}=(-1)^{n+k}s\cos
\dfrac{\theta}{2}+O\left(n^{-1/4}\log^{1/2}n+
\left(\dfrac{\left|k\right|}{n}\right)^{1/2}\right).
\end{equation}
\end{theorem}
\textit{Proof of Theorem~\ref{t:NullApp}.} The main idea of the proof
is to derive  an equation with a functional parameter $\phi$ for functions
$\psi_{k}^{(n)}$ of \eqref{d:Psi_k} from the determinant formulas.
Then, choosing appropriate parameter $\phi$,
we obtain the equation for the Verblunsky coefficients.
Define a new
eigenvalue distribution
\begin{equation}\label{rho_k_n}
p_{k,n} \left( \lambda_1, \ldots, \lambda_k\right)= Q_{k,n}^{-1}
\prodd_{1 \leq j < m \leq k} \left|e^{i \lambda_j}-e^{i
\lambda_m}\right|^2 \exp\left\{-n \sumd_{l=1}^{k}
V\left(\cos \lambda_l\right)\right\},
\end{equation}
which differs from \eqref{d:PDef} by the number of variables ($k$ instead of $n$).
Let
\begin{equation}\label{rho_k_n_1}
\begin{array}{c}
\widetilde{\rho}_{k,n} \left(\lambda\right)=\intd p_{k,n} \left(
\lambda,\lambda_2, \ldots, \lambda_k\right) d \lambda_2 \ldots d
\lambda_k,
\\%
\widetilde{\rho}_{k,n} \left(\lambda, \mu\right)=\intd p_{k,n}
\left( \lambda,\mu, \lambda_3, \ldots, \lambda_k\right) d \lambda_3
\ldots d \lambda_k,
\end{array}
\end{equation}
be the first and the second marginal densities of \eqref{rho_k_n}.
By the standard argument (see [\onlinecite{Me:91}]) we have
\begin{equation}\label{rho_k_n-K}
\begin{array}{c}
\widetilde{\rho}_{k,n} \left(\lambda\right)=k^{-1} K_{k,n}
\left(\lambda, \lambda\right),
\\%
\widetilde{\rho}_{k,n} \left(\lambda, \mu\right) =
\dfrac{1}{k\left(k-1\right)} \left[K_{k,n}\left( \lambda,
\lambda\right) K_{k,n}\left( \mu, \mu\right) - \left|K_{k,n}\left(
\lambda, \mu\right)\right|^2\right],
\end{array}
\end{equation}
\noindent%
where
$K_{k,n}$ defined in \eqref{d:K_k_n}.
We will use also the notation
\begin{equation}\label{rh_k_n}
\rho_{k,n} \left(\lambda\right) \equiv \dfrac 1n
K_{k,n}\left(\lambda,\lambda\right)=\dfrac kn \widetilde{\rho}_{k,n}
\left(\lambda\right).
\end{equation}
Let $\phi$ be any $2\pi$~-periodic twice differentiable function.
Then integrating by parts and using that $\dfrac{d}{d\lambda}\left|e^{i\lambda}-e^{i\mu}\right|^2=
\cot{\dfrac{\lambda-\mu}{2}}\left|e^{i\lambda}-e^{i\mu}\right|^2$, we get
\begin{multline*}
\intd_{-\pi}^{\pi} \left(V \left(\cos \lambda\right)\right)' \widetilde{\rho}_{k+n,n}
\left(\lambda\right) \phi \left(\lambda\right) d \lambda = n^{-1}
\intd_{-\pi}^{\pi} \widetilde{\rho}_{k+n,n} \left(\lambda\right)
\phi' \left(\lambda\right) d \lambda
\\%
+\dfrac{n+k-1}{n} \intd_{-\pi}^{\pi} \widetilde{\rho}_{k+n,n}
\left(\lambda, \mu\right) \dfrac{\phi \left(\lambda\right)}{\tan
\left(\lambda-\mu\right)/2} d \lambda d \mu.
\end{multline*}
Combining the above relation with the symmetry property
$\widetilde{\rho}_{k+n,n} \left(\lambda,
\mu\right) = \widetilde{\rho}_{k+n,n} \left(\mu, \lambda\right)$, we
obtain
\begin{multline*}
\intd_{-\pi}^{\pi} \left(V \left(\cos \lambda\right)\right)' \widetilde{\rho}_{k+n,n}
\left(\lambda\right) \phi \left(\lambda\right) d \lambda = n^{-1}
\intd_{-\pi}^{\pi} \widetilde{\rho}_{k+n,n} \left(\lambda\right)
\phi' \left(\lambda\right) d \lambda
\\%
+\dfrac{n+k-1}{2n} \intd_{-\pi}^{\pi} \widetilde{\rho}_{k+n,n}
\left(\lambda, \mu\right) \dfrac{\phi \left(\lambda\right) - \phi
\left(\mu\right)}{\tan\left(\lambda-\mu\right)/2} d \lambda d \mu.
\end{multline*}
Using \eqref{rho_k_n-K}, \eqref{rh_k_n} and the equality
\begin{equation*}
\intd_{-\pi}^{\pi} \left|K_{k+n,n} \left(\lambda,
\mu\right)\right|^2 d \mu = K_{k+n,n} \left(\lambda, \lambda\right),
\end{equation*}
we obtain from the above
\begin{eqnarray}
\intd_{-\pi}^{\pi} \dfrac{\phi \left(\lambda\right) - \phi
\left(\mu\right)}{2 \tan\left(\lambda-\mu\right)/2} \rho_{k+n,n}
\left(\lambda\right)
\rho_{k+n,n} \left(\mu\right)  d \lambda d \mu & \notag
\\
-%
\intd_{-\pi}^{\pi} \left(V \left(\cos \lambda\right)\right)' & \rho_{k+n,n}
\left(\lambda\right) \phi \left(\lambda\right) d \lambda +
\delta_{k,n} \left[\phi\right]=0,\label{eq_rho_3}
\end{eqnarray}
where
\begin{equation}\label{delta_0}
\delta_{k,n} \left[\phi\right]:= \dfrac{1}{2n^2}%
\intd_{-\pi}^{\pi} \left[\phi'\left(\lambda\right)+\phi'
\left(\lambda\right) - \dfrac{\phi \left(\lambda\right) - \phi
\left(\mu\right)}{\tan\left(\lambda-\mu\right)/2}\right]%
\left|K_{k+n,n} \left(\lambda, \mu\right)\right|^2 d \lambda d \mu.
\end{equation}
We replace in \eqref{eq_rho_3} $k$ by $k-1$, subtract the new
relation from \eqref{eq_rho_3}, and multiply the result by $n$. Then
we have
\begin{multline}\label{eq_rho_4}
\intd_{-\pi}^{\pi} \dfrac{\phi \left(\lambda\right) - \phi
\left(\mu\right)}{\tan\left(\lambda-\mu\right)/2} \rho\left(\mu\right)
\left| \psi^{\left(n\right)}_{k+n} \left(\lambda\right)\right|^2  d \lambda d \mu-%
\intd_{-\pi}^{\pi} \left(V \left(\cos \lambda\right)\right)' \phi
\left(\lambda\right)\left|\psi^{\left(n\right)}_{k+n}
\left(\lambda\right)\right|^2  d \lambda
\\
+ \delta^{\left(1\right)}_{k,n}
\left[\phi\right]+\delta^{\left(2\right)}_{k,n} \left[\phi\right]=0,
\end{multline}
where
\begin{multline*}
\delta^{\left(1\right)}_{k,n} \left[\phi\right]:=  \dfrac{1}{n}\intd_{-\pi}^{\pi}
\left[\phi'\left(\lambda\right)+\phi' \left(\mu\right) - \dfrac{\phi
\left(\lambda\right) - \phi \left(\mu\right)}{\tan\left(\lambda-\mu\right)/2}\right]%
\\
\cdot \Re\{K_{k+n-1,n} \left(\lambda, \mu\right)
\overline{\psi^{\left(n\right)}_{k+n} \left(\lambda\right)}
\psi^{\left(n\right)}_{k+n} \left(\mu\right)\} d \lambda d \mu,
\end{multline*}
\begin{multline*}
\delta^{\left(2\right)}_{k,n} \left[\phi\right]:= \intd_{-\pi}^{\pi} \dfrac{\phi
\left(\lambda\right) - \phi \left(\mu\right)}{\tan\left(\lambda-\mu\right)/2}
\left(\rho_{n+k-1,n} \left(\mu\right) -
\rho\left(\mu\right)\right) \left| \psi^{\left(n\right)}_{k+n}
\left(\lambda\right)\right|^2 d \lambda d \mu
\\
+ \dfrac{1}{n}
\intd \phi'\left(\lambda \right)\left| \psi^{\left(n\right)}_{k+n}
\left(\lambda\right)\right|^2 d \lambda.
\end{multline*}
To estimate $\delta^{\left(1\right)}_{k,n} \left[\phi\right]$ we note that
\begin{equation*}
\left|\phi'\left(\lambda\right)+\phi' \left(\mu\right) - \dfrac{\phi
\left(\lambda\right) - \phi \left(\mu\right)}{\tan\left(\lambda-\mu\right)/2}\right|
\leq C \left| e^{i\lambda}-e^{i\mu}
\right| \left(\|\phi'\|_{\infty}+\|\phi''\|_{\infty}\right).
\end{equation*}
and use also the bound (see e.g. [\onlinecite{Popl:08}])
\begin{equation*}
\intd \left|K_{k+n-1,n} \left(\lambda, \mu\right)\right|^2 %
\left|e^{i\lambda}-e^{i\mu} \right|^2 d \lambda d \mu \leq C.
\end{equation*}
Combining it with the Schwartz inequality, we have
\begin{multline}\label{e:del_1}
\left|\delta^{\left(1\right)}_{k,n} \left[\phi\right]\right| \leq \dfrac{C}{n}
\left(\|\phi'\|_{\infty}+\|\phi''\|_{\infty}\right) %
\left(%
\intd \left|K_{k+n-1,n} \left(\lambda, \mu\right)\right|^2 %
\left|e^{i\lambda}-e^{i\mu} \right|^2 d \lambda d \mu
\right)^{1/2}%
\\
\cdot%
\intd
\left|%
\psi^{\left(n\right)}_{k+n} \left(\lambda\right)
\psi^{\left(n\right)}_{k+n} \left(\mu\right)
\right|^2%
d \lambda d \mu \leq \dfrac{C}{n}
\left(\|\phi'\|_{\infty}+\|\phi''\|_{\infty}\right).
\end{multline}
For  $\delta^{\left(2\right)}_{k,n} \left[\phi\right]$ we use the following result of
[\onlinecite{Kol:97}]. Let $\rho_n:=p_1^{(n)}$ be the first marginal density of \eqref{d:PDef}.
Then $\rho_n$  converges weakly to the equilibrium density $\rho$ and
for any $\phi \in H^1 \left(-\pi,\pi\right)$
\begin{equation}\label{e:WConv}
\left|%
\intd\phi\left(\lambda\right)\rho_n\left(\lambda\right)%
\,d \lambda%
-%
\intd\phi\left(\lambda\right)\rho
\left(\lambda\right)%
\,d \lambda%
\right|%
\leq%
C \left\| \phi \right\|^{1/2}_{2} \left\| \phi' \right\|^{1/2}_{2}
n^{-1/2} \ln^{1/2} n,
\end{equation}
where $\left\|\cdot\right\|_2$ denotes $L_2$ norm on $[-\pi,\pi]$.
Since the function $\dfrac{\phi \left(\lambda\right)
- \phi \left(\mu\right)}{\tan\left(\lambda-\mu\right)/2}$ has a bounded
derivative with respect to $\mu$, we obtain
\begin{eqnarray}
\left|\delta^{\left(2\right)}_{k,n} \left[\phi\right]\right| &\leq&%
\left|\delta^{\left(2\right)}_{1,n} \left[\phi\right]\right|+C
\dfrac{\left|k\right|}{n}\|\phi'\|_{\infty} \notag
\\
&\leq C& %
\left( %
\left\|\phi' \right\|^{1/2}_{\infty} %
\left\| \phi''\right\|^{1/2}_{\infty}%
n^{-1/2} \log^{1/2} n+%
\dfrac{\left|k\right|}{n}\|\phi'\|_{\infty}
\right).\label{e:del_2}
\end{eqnarray}
We split integral in \eqref{eq_rho_4}  into two parts.
Using \eqref{e:V_Rho} for $\lambda \in \sigma$, we obtain
\begin{multline}\label{e:eq_rho_5}
v.p. \intd_{\sigma} d \lambda \intd_{\sigma} d \mu \phi
\left(\mu\right)\cot \dfrac{\lambda-\mu}{2} \rho\left(\mu\right)
\left| \psi^{\left(n\right)}_{k+n} \left(\lambda\right)\right|^2
\\
-\intd_{\sigma^c} d\lambda
\left|\psi^{\left(n\right)}_{k+n}
\left(\lambda\right)\right|^2
\left(
\intd_{\sigma} d\mu
\rho\left(\mu\right)
\dfrac{\phi
\left(\lambda\right) - \phi \left(\mu\right)}{\tan\left(\lambda-\mu\right)/2}
-
\phi\left(\lambda\right)
\left(V \left(\cos \lambda\right)\right)'
\right)
\\
= \delta^{\left(1\right)}_{k,n} \left[\phi\right]+\delta^{\left(2\right)}_{k,n}
\left[\phi\right].
\end{multline}
Equation \eqref{e:V_Rho} is valid only for $\lambda \in \sigma$. Below
we need to extend this relation on some
neighborhood of $\sigma$. Using \eqref{e:Rho_a}, we obtain the following  representation.
\begin{lemma}\label{l:V_R}
Under the conditions of Theorem \ref{t:Main}, for
$\varepsilon$ defined in C3 and
$\lambda \in \sigma_{\varepsilon}= \left[-\theta-\varepsilon,\theta+\varepsilon\right]$,
$V$ can be represented in the
form
\begin{multline}\label{e:V_R}
\left(V \left(\cos \lambda\right)\right)'-\dfrac{1}{4\pi^2} \intd_{-\theta}^{\theta}
\cot \dfrac{\lambda-\mu}{2} P\left(\mu\right) \chi \left(\mu\right)
d\mu
\\
=\dfrac{\mathrm{sign} \lambda}{2\pi}\left( P\left(\lambda\right)
\chi \left(\lambda\right)+O\left(\chi^2 \left(\lambda\right)\right)\right)\mathbf{1}_{\sigma_{\varepsilon} \setminus \sigma}, \, \text{for} \, \lambda \to \pm \theta.
\end{multline}
\end{lemma}
Proofs of this lemma and other auxiliary results can be found in Section~\ref{s:Auxillary} of the paper.  For $\lambda \in \sigma_{\varepsilon}^c$ we use the exponential bounds for the functions $\psi_k^{(n)}$.
These bounds can be proved analogously to the corresponding result of [\onlinecite{Pa-Sh:07}].
\begin{proposition}\label{pr:Psi_n+k}
Let potential $V$ satisfy conditions C1-C3. Than there exist
constants $C,\, C_1 \, >0$ and $\varepsilon_1>0$
such that for any integer $k$ satisfying inequality $|k|<\varepsilon_1 n$
we have
\begin{equation}\label{e:Phi_eps}
\intd_{\sigma_{\varepsilon_{n,k}}} \left| \psi^{\left(n\right)}_{k+n}
\left(\lambda\right)\right|^2 d \lambda \leq e^{-C n^{1/2}\log n},
\quad\varepsilon_{n,k} = C_1\left(n^{-1/2}\log n + \left|k\right|/n\right).
\end{equation}
\end{proposition}
Combining \eqref{e:V_R}, \eqref{e:eq_rho_5}, \eqref{e:del_1}, \eqref{e:del_2}
with \eqref{e:Phi_eps}, we obtain finally the integral equation
\begin{multline}\label{e:eq_rho_a}
v.p. \intd_{\sigma_{\varepsilon_{n,k}}} d \lambda \intd_{-\theta}^{\theta} d \mu \phi
\left(\mu\right)\cot \dfrac{\lambda-\mu}{2} \rho\left(\mu\right)
\left| \psi^{\left(n\right)}_{k+n} \left(\lambda\right)\right|^2
\\
+\dfrac{1}{2\pi} \intd_{\sigma_{\varepsilon_{n,k}}\setminus \sigma}\hbox{sign} \lambda \phi
\left(\lambda\right) P\left(\lambda\right) \chi\left(\lambda\right) \left|\psi^{\left(n\right)}_{k+n}
\left(\lambda\right)\right|^2 d \lambda
\\
= \left(\left\|\phi\right\|_{\infty}+\left\|\phi'\right\|_{\infty}
+\left\|\phi''\right\|_{\infty}\right)O\left(\varepsilon_{n,k}\right)
\end{multline}
with $\varepsilon_{n,k}$ of (\ref{e:Phi_eps}).
\begin{remark}
If the support of the equilibrium measure is the whole interval $\left[-\pi,\pi\right]$, then
 we can obtain in a similar way the integral equation for the equilibrium density
\begin{equation}\label{e:eq_rho_w}
v.p. \intd_{-\pi}^{\pi} \phi \left(\mu\right)\cot
\dfrac{\lambda-\mu}{2} \rho\left(\mu\right) \left|
\psi^{\left(n\right)}_{k+n} \left(\lambda\right)\right|^2 d \lambda
d \mu = \left(\left\|\phi'\right\|_{\infty}
+\left\|\phi''\right\|_{\infty}\right)O\left(n^{-1/2}\log{n}+\dfrac{|k|}{n}\right).
\end{equation}
\end{remark}
It is easy to see that $P\left(\lambda\right)$ of \eqref{d:Sqrt} is a twice
differentiable strictly positive function on the interval
$\left[-\theta,\theta\right]$. Taking in \eqref{e:eq_rho_a}
\begin{equation*}
\phi \left(\mu\right) = P^{-1}\left(\mu\right) \cos \dfrac{\mu}{2}
\cot \dfrac{z-\mu}{2},
\end{equation*}
we obtain
\begin{multline}\label{e:arc}
\intd_{\sigma_{\varepsilon_n}}\left| \psi^{\left(n\right)}_{k+n}
\left(\lambda\right)\right|^2 f\left(z,\lambda, \theta\right)%
d \lambda +
\\
+2 \pi \intd_{_{\sigma_{\varepsilon_n}} \backslash \sigma}%
\hbox{sign} \lambda \cot \dfrac{z-\lambda}{2}\cos \dfrac{\lambda}{2}
\chi \left(\lambda\right) \left| \psi^{\left(n\right)}_{k+n}
\left(\lambda\right)\right|^2 d\lambda
=O\left(\varepsilon_{n,k}
\left|\mathrm{dist}\left(z,\sigma\right)\right|^{-3}\right),
\end{multline}
where
\begin{equation}\label{d:fzlt}
f\left(z,\lambda, \theta\right) = v.p.\intd_{-\theta}^{\theta} \cot
\dfrac{\lambda-\mu}{2} \cot \dfrac{z-\mu}{2} \chi \left(\mu\right)
\cos \dfrac{\mu}{2} d\mu.
\end{equation}
Computing $f\left(z,\lambda,\theta\right)$
we obtain
\begin{multline*}
f\left(z,\lambda, \theta\right) = v.p.\intd_{-\theta}^{\theta}
\dfrac{\sin \mu + \sin \lambda}{\cos \mu - \cos \lambda}%
\dfrac{\sin \mu + \sin z}{\cos \mu - \cos z}%
\chi \left(\mu\right) \cos \dfrac{\mu}{2} d\mu=
\\
=v.p.\intd_{-\theta}^{\theta}
\dfrac{\sin^2 \mu + \sin \lambda \sin z}%
{\left(\cos \mu - \cos \lambda\right)\left(\cos \mu - \cos z\right)}%
\chi \left(\mu\right) \cos \dfrac{\mu}{2} d\mu=
\\
=v.p.\intd_{-\theta}^{\theta} %
\left[%
-1+\cot \dfrac{z-\lambda}{2}%
\left(%
\dfrac{\sin \lambda}{\cos\mu-\cos\lambda}-
\dfrac{\sin z}{\cos\mu-\cos z}%
\right)%
\right]%
\chi \left(\mu\right) \cos \dfrac{\mu}{2} d\mu=
\\
=I_1+\cot \dfrac{z-\lambda}{2} \left(\sin \lambda I_2 \left(\lambda\right)-\sin z
I_2 \left(z\right)\right),
\end{multline*}
where
\begin{equation*}
I_1= - \intd_{-\theta}^{\theta} %
\chi \left(\mu\right) \cos \dfrac{\mu}{2} d\mu = -\pi \sqrt{2}
\sin^2 \dfrac{\theta}{2}, \quad I_2\left(z\right)=v.p. \intd_{-\theta}^{\theta} %
\dfrac{\chi \left( \mu \right)}{\cos \mu - \cos z} \cos
\dfrac{\mu}{2} d\mu.
\end{equation*}
To compute $I_2\left(z\right)$ we use well known relations (see [\onlinecite{Mu:53}])
\begin{equation}\label{e:Int_sq}
Sq_a\left(z\right):=\dfrac{1}{\pi}v.p.\intd_{-a}^{a} \dfrac{dt}{\left(z-t\right) \sqrt{a^2-t^2}}
=
\left\{
\begin{array}{cc}
1/\sqrt{z^2-a^2}, & \mbox{for} \, z \notin \left[-a,a\right],\\
0, & z \in \left(-a,a\right),
\end{array}
\right.
\end{equation}
where $1/\sqrt{z^2-a^2}$ is defined by the asymptotic $\sim 1/z, \, z \to \infty$. Then we have for $z\neq \pm \theta$
\begin{equation*}
I_2\left(z\right)=
\sqrt{2} \intd_{-1}^{1} \dfrac{\sqrt{1-t^2}dt}{\sin^2\left(z/2\right)/\sin^2\left(\theta/2\right)-t^2}
=
\sqrt{2}\pi\left(1-\dfrac{\sqrt{\sin^2z/2-\sin^2\theta/2}}{\sin z/2}\mathbf{1}_{\sigma^c}\right).
\end{equation*}
Now, using the above results,  we obtain from \eqref{e:eq_rho_a}
\begin{multline}\label{e:Psi_eq}
\intd_{\sigma_{\varepsilon_{n,k}}} \left|
\psi^{\left(n\right)}_{k+n}\left(\lambda\right)\right|^2
\left[-\sin^2 \dfrac{\theta}{2}+\cot \dfrac{z-\lambda}{2}\left(\sin
\lambda -\sin z\right) \right] d\lambda
\\
+ \intd_{\sigma_{\varepsilon_{n,k}}} \left|
\psi^{\left(n\right)}_{k+n}\left(\lambda\right)\right|^2 \sin z
\dfrac{\sqrt{\sin^2z/2-\sin^2\theta/2}}{\sin z/2}\cot \dfrac{\lambda-z}{2} d\lambda
=O\left(\varepsilon_{n,k}
\left|\mathrm{dist}\left(z,\sigma\right)\right|^{-3}\right).
\end{multline}
Denote $R^{\left(n\right)}$ the analog of the resolvent for the matrix $C^{\left(n\right)}$:
\begin{equation}\label{d:Resolvent}
R^{\left(n\right)}=i\dfrac{C^{\left(n\right)}+e^{iz}}{C^{\left(n\right)}-e^{iz}}.
\end{equation}
Then by the spectral theorem
\begin{eqnarray*}
&
R^{\left(n\right)}_{n+k,n+k} \left(z\right) = \intd \left|
\psi^{\left(n\right)}_{k+n}\left(\lambda\right)\right|^2
\cot \dfrac{\lambda-z}{2} d\lambda,
\\
&
C^{\left(n\right)}_{n+k,n+k}+\cos z =
\intd \left|
\psi^{\left(n\right)}_{k+n}\left(\lambda\right)\right|^2
\cot \dfrac{\lambda-z}{2} \left(\sin \lambda - \sin z\right)d\lambda
.
\end{eqnarray*}
The above relations combined with \eqref{e:Phi_eps} allow us to derive from \eqref{e:Psi_eq} that
\begin{equation}\label{e:Rez_a}
R^{\left(n\right)}_{n+k,n+k} \left(z\right)= %
\dfrac{C^{\left(n\right)}_{n+k,n+k}+\cos z+\sin^2
\theta/2+O\left(\varepsilon_{n,k}\left|\mathrm{dist}\left(z,\sigma\right)
\right|^{-3}\right)}{2\cos \left(z/2\right) \, \sqrt{\sin^2z/2-\sin^2\theta/2}}.
\end{equation}
Integrating  the product $e^{iz}R^{\left(n\right)}_{n+k,n+k}
\left(z\right)$ over the contour
$\mathfrak{L}_{\delta}=\left\{\left.z\right| \mathrm{dist}
\left(z,\sigma\right)=\delta\right\}$ for some $\delta$ we
obtain from \eqref{e:Rez_a}
\begin{equation}\label{e:Int_a}
\oint_{\mathfrak{L}_{\delta}}
e^{iz}R^{\left(n\right)}_{n+k,n+k}\left(z\right) dz =
\oint_{\mathfrak{L}_{\delta}}e^{iz}
\dfrac{C^{\left(n\right)}_{n+k,n+k}+\cos z+\sin^2
\theta/2}{2\cos \left(z/2\right) \, \sqrt{\sin^2z/2-\sin^2\theta/2}}dz
+O\left(\varepsilon_{n,k}\delta^{-7/2}\right).
\end{equation}
The Cauchy theorem and the spectral theorem yield
\begin{equation}\label{e:Exp_Cot}
\oint_{\mathfrak{L}_{\delta}} e^{iz}R_{n+k,n+k} dz = - 4 i \pi
C^{\left(n\right)}_{n+k,n+k}.
\end{equation}
Since the integrand in the r.h.s. of \eqref{e:Int_a} is analytic outside $\sigma$
the integral does not depend on $\delta$, therefore we
can compute this integral with $\delta
\rightarrow 0$.
\begin{eqnarray*}
I\left(C^{\left(n\right)}_{n+k,n+k}\right)&:=\lim_{\delta \rightarrow 0}\oint_{\mathfrak{L}_{\delta}}e^{iz}
\dfrac{C^{\left(n\right)}_{n+k,n+k}+\cos z+\sin^2
\theta/2}{2\cos \left(z/2\right) \, \sqrt{\sin^2z/2-\sin^2\theta/2}}dz &
\\
&= -i\intd_{-\theta}^{\theta} \cos \mu
\dfrac{C^{\left(n\right)}_{n+k,n+k}+\cos \mu+\sin^2
\theta/2}{\cos \left(\mu/2\right) \sqrt{\sin^2
\theta/2-\sin^2 \mu/2}}d\mu.
\end{eqnarray*}
Then, using the change of the variable $\sin \mu/2 = t$ with \eqref{e:Int_sq} we get
\begin{equation}\label{e:C_a}
I\left(C^{\left(n\right)}_{n+k,n+k}\right)=-4i\pi
C^{\left(n\right)}_{n+k,n+k} - \dfrac{2i\pi}{\cos
\theta/2}\left(C^{\left(n\right)}_{n+k,n+k}-\cos^2
\theta/2\right).
\end{equation}
Combining \eqref{e:arc}, \eqref{e:Exp_Cot} and \eqref{e:C_a}, we
have
\begin{equation}\label{e:as_a}
C^{\left(n\right)}_{n+k,n+k}=\cos^2
\dfrac{\theta}{2}+O\left(\varepsilon_{n,k}\right).
\end{equation}
Then in view of \eqref{e:Rez_a} we obtain for the "resolvent"
\begin{equation}\label{e:Res_as}
R^{\left(n\right)}_{n+k,n+k} \left(z\right)=
\dfrac{1+\cos z +
O\left(\varepsilon_{n,k}\left|\mathrm{dist}\left(z,\sigma\right)
\right|^{-3}\right)}{2\cos z/2 \, \sqrt{\sin^2z/2-\sin^2\theta/2}}.
\end{equation}
Similarly, integrating the product
$e^{2iz}R^{\left(n\right)}_{n+k,n+k} \left(z\right)$ over the contour
$\mathfrak{L}_{\delta}$ and taking $\delta \rightarrow 0$, we get
\begin{multline}\label{e:C^2_a}
-4i \pi \left(C^{\left(n\right)}\right)^2_{n+k,n+k}=-i
\intd_{-\theta}^{\theta} \dfrac{2\cos\mu/2 \, \, \cos2\mu}
{\sqrt{\sin^2 \theta/2-\sin^2
\mu/2}}d\mu+O\left(\varepsilon_{n,k}\right)
\\
= -4i \pi\left(3\sin^4 \dfrac{\theta}{2}-4\sin^2\dfrac{\theta}{2}+1
+O\left(\varepsilon_{n,k}\right)\right).
\end{multline}
From \eqref{m:CMV} we have
\begin{multline}\label{e:C^2}
\left(C^{\left(n\right)}\right)^2_{m,m}=%
C^{\left(n\right)}_{m,m-1} C^{\left(n\right)}_{m-1,m}+
C^{\left(n\right)}_{m,m} C^{\left(n\right)}_{m,m}+
C^{\left(n\right)}_{m,m+1} C^{\left(n\right)}_{m+1,m}
\\
=C^{\left(n\right)}_{m,m} C^{\left(n\right)}_{m-1,m-1}+
C^{\left(n\right)}_{m,m} C^{\left(n\right)}_{m,m}+
C^{\left(n\right)}_{m,m} C^{\left(n\right)}_{m+1,m+1}-
\alpha_{m-1}^{\left(n\right)}\alpha_{m+1}^{\left(n\right)}-
\alpha_{m}^{\left(n\right)}\alpha_{m+2}^{\left(n\right)}.
\end{multline}
Taking into account \eqref{e:as_a} and \eqref{e:C^2_a}, we obtain
from \eqref{e:C^2}
\begin{equation}\label{e:alpha_a}
\left\{
\begin{array}{c}
\alpha_{n+k-1}^{\left(n\right)}\alpha_{n+k+1}^{\left(n\right)} +
\alpha_{n+k}^{\left(n\right)}\alpha_{n+k+2}^{\left(n\right)} =
2\cos^2 \dfrac{\theta}{2} +O\left(\varepsilon_{n,k}\right)
\\
\alpha_{n+k-1}^{\left(n\right)}\alpha_{n+k}^{\left(n\right)} =
-\cos^2 \dfrac{\theta}{2} +O\left(\varepsilon_{n,k}\right)
\end{array}
\right.
.
\end{equation}
To solve the system  we multiply the first equation in \eqref{e:alpha_a} by
$\alpha_{n+k}^{\left(n\right)}\alpha_{n+k+1}^{\left(n\right)}$.
By the definition of the Verblunsky coefficients $\left|\alpha_{j}^{\left(n\right)}\right| \leq 1$, hence
using the second equation of \eqref{e:alpha_a} we obtain
\begin{equation*}
\left(\alpha_{n+k+1}^{\left(n\right)}\right)^2 +
\left(\alpha_{n+k}^{\left(n\right)}\right)^2
=
2\cos^2 \dfrac{\theta}{2} +O\left(\varepsilon_{n,k}\right).
\end{equation*}
Finally, combining the above relations, we obtain \eqref{e:NullApp}.
\begin{flushright}
$\blacksquare$
\end{flushright}
\begin{remark}\label{r:Asym_w}
If the support of the equilibrium measure is the whole circle we can use function
$\phi\left(\mu\right)=\cot \dfrac{z-\mu}{2}
\rho^{-1}\left(\mu\right)$ in \eqref{e:eq_rho_w}
and obtain similarly to \eqref{e:as_a}
\begin{equation}\label{e:Alpha_Main_w}
C^{\left(n\right)}_{n+k,n+k}=O\left(n^{-1/2}\log n+
\dfrac{\left|k\right|}{n}\right).
\end{equation}
\end{remark}
\textit{Proof of Theorem~\ref{t:Main}.}
Let
\begin{equation}\label{e:x_def}
\alpha_{n+m}^{\left(n\right)}=\left(-1\right)^{n+m} s
\cos \left(\dfrac{\theta}{2}+x^{\left(n\right)}_{n+m}\right), \quad
\rho_{n+m}^{\left(n\right)}=\sin \left(\dfrac{\theta}{2}+x^{\left(n\right)}_{n+m}\right),
\end{equation}
where $s=\pm1$ and $x_{n+m} \in [-\pi,\pi)$. We have already a priori approximate \eqref{e:NullApp}
\begin{equation}\label{e:x_asymp}
x^{\left(n\right)}_{n+m}=\underline{O} \left(n^{-1/4}  \log^{1/2} n + \left|\dfrac{m}{n}\right|^{1/2}\right).
\end{equation}
In this part of the paper we use the string
equation for unitary matrix models \eqref{e:Str_eq}. We
consider these equations as a system of nonlinear equations for
$x^{\left(n\right)}_{n+m}$.
Denote
\begin{equation}\label{e:M_def}
M^{\left(n\right)}=\dfrac{C^{\left(n\right)}+C^{\left(n\right)*}}{2}, \quad
L^{\left(n\right)}=\dfrac{C^{\left(n\right)}-C^{\left(n\right)*}}{2}.
\end{equation}
Equation \eqref{e:Str_eq} can be rewritten as
\begin{equation}\label{e:str_eq.m}
\left\{L^{\left(n\right)} V' \left(M^{\left(n\right)}\right)\right\}_{n+k,n+k-1}=
 \left(-1\right)^{n+k} \dfrac{n+k}{n} \dfrac{\alpha_{n+k}^{\left(n\right)}}
{\rho_{n+k}^{\left(n\right)}}.
\end{equation}
To simplify notations we set
\begin{equation}\label{e:M_tilde_def}
\widetilde{M}_{j,k} = M^{\left(n\right)}_{n+j,n+k},
\quad
\widetilde{L}_{j,k} = L^{\left(n\right)}_{n+j,n+k},
\quad
\widetilde{\alpha}_{k}=\alpha^{\left(n\right)}_{n+k},
\quad
\widetilde{x}_{k}=x^{\left(n\right)}_{n+k},
\end{equation}
\begin{equation}\label{e:abc}
a=\cos^2 \dfrac{\theta}{2}, \quad
b=\cos \dfrac{\theta}{2} \sin \dfrac{\theta}{2}, \quad
c=\sin^2 \dfrac{\theta}{2}.
\end{equation}
and define $M^{\star}$ as a double-infinite matrix with entries
\begin{equation}\label{e:M_jk}
M^{\star}_{j,k}=\delta_{j,k} a + \dfrac{1}{2}\left(\delta_{j-2,k}+\delta_{j,k-2}\right)c.
\end{equation}
The spectrum and the generalized eigenvectors of the matrix can be parameterized as
\begin{equation}\label{e:M0_sp}
M^{\star} \sim
\left\{
a+c\cos 2\lambda, \, \lambda \in \left[-\pi,\pi\right];
v_{k}=\dfrac{1}{\sqrt{2\pi}}e^{ik\lambda}
\right\}.
\end{equation}
From now  we consider matrix $\widetilde{M}$ as a double-infinite one and redefine it for
$j,k<-n$ equal to $M^{\star}$.
It is convenient for us to extend the function $V'$ from $\left[-1,1\right]$
to $\mathbb{R}$ in such a way to obtain the function from $\mathcal{L}_2 \left(\mathbb{R}\right)$ with
the  third derivative bounded in $\mathcal{L}_2 \left(\mathbb{R}\right)$.
Denoting $\widehat{V'}$ the Fourier transform of the extension we have for $x\in \left[0,2\pi\right]$
\begin{equation}\label{e:v_m_def}
V'\left(x\right) = \dfrac{1}{2\pi} \intd_{-\infty}^{\infty} \widehat{V'} \left(\xi\right) e^{ix\xi} d\xi.
\end{equation}
Using \eqref{e:v_m_def}, $V'\left(M^{\left(n\right)}\right)$ can be represented
\begin{equation}\label{e:V`_M}
V' \left( \widetilde{M} \right)_{i,j}
= \dfrac{1}{2\pi}\intd_{-\infty}^{\infty}  \widehat{V'} \left(\xi\right)
\left\{e^{i\xi \widetilde{M}}\right\}_{i,j}d\xi
.
\end{equation}
Applying the Duhamel formula twice to matrices $\widetilde{M}$ and $M^*$, we obtain
\begin{multline}\label{e:m_duh}
e^{i\xi \widetilde{M}}=
e^{i\xi M^{\star}} + i\xi
\intd_0^1 e^{i\xi M^{\star}t} \left(\delta M\right)e^{i\xi M^{\star} \left(1-t\right)} dt
\\
-
\xi^2
\intd_{0}^{1} dt \intd_{0}^{1-t} ds e^{i\xi M^{\star} t} \left(\delta M\right) e^{i\xi \widetilde{M}s} \left(\delta M\right) e^{i\xi M^{\star} (1-t-s)},\quad
\text{for} \, \, \delta M = \widetilde{M}-M^{\star}.
\end{multline}
Denote for $d=0,\pm1,\pm2$
\begin{equation}\label{e:g_j_def}
\begin{array}{ll}
\hat{f}\left(\xi\right)   =  \left\{e^{i\xi \widetilde{M}}\right\}_{k+d,k-1},&
\hat{f}^0\left(\xi\right) =  \left\{e^{i\xi M^*}\right\}_{k+d,k-1}, \\
\hat{f}^1\left(\xi\right) =  \left\{i\xi
\intd_0^1 e^{i\xi M^*t} \left(\delta M\right)e^{i\xi M^*\left(1-t\right)} dt \right\}_{k+d,k-1},&
\widehat{r}\left(\xi;d\right)   =  \hat{f}\left(\xi\right) - \hat{f}^0\left(\xi\right)-\hat{f}^1\left(\xi\right).
\end{array}
\end{equation}
Now we find the first two terms in \eqref{e:m_duh}.
\begin{multline}\label{e:V_M}
\hat{f}^0\left( \xi \right)=
\left\{e^{i\xi M^*}\right\}_{k+d,k-1}=
\dfrac{1}{2\pi}\intd_{-\pi}^{\pi} e^{i \xi \left(a+c\cos2\phi\right)}
e^{i\left(k+d\right)\phi} e^{-i\left(k-1\right)\phi} d\phi
\\
=
\dfrac{1}{2\pi}\intd_{-\pi}^{\pi} e^{i \xi \left(a+c\cos2\phi\right)}
\cos\left(d+1\right)\phi d\phi
.
\end{multline}
Let $\left\{v_k\right\}$ be the coefficients of the Fourier series for the function $V'\left(a+c\cos2\phi\right)$,
i.e.,
\begin{equation}\label{d:v_k}
v_k=\dfrac{1}{\pi \left(1+\delta_{k,0}\right)}\intd_{-\pi}^{\pi}V'\left(a+c\cos2\phi\right) \cos k\phi d\phi.
\end{equation}
Note that $v_{2l+1}=0$, for any integer $l$. By the spectral theorem we obtain
\begin{multline}\label{e:null_p}
\left\{
\widetilde{L} V'\left(M^{\star}\right)
\right\}_{k,k-1}=
\sumd_{d=0,\pm 1,\pm 2} \widetilde{L}_{k,k+d} \dfrac{1}{2\pi} \intd_{-\pi}^{\pi}
V' \left(a+c\cos2\phi\right) \cos\left(d+1\right)\phi d\phi
\\
=
\dfrac{\widetilde{L}_{k,k+1}}{c}
+\left(\widetilde{L}_{k,k+1}+\widetilde{L}_{k,k-1}\right)
v_0
,
\end{multline}
where we have used the relation
\begin{proposition}\label{p:v_1_v_0}
Under conditions C1-C3 for $v_k$ defined in \eqref{d:v_k}
$$
v_2=2v_0+\dfrac{2}{c}.
$$
\end{proposition}
For the second term in \eqref{e:m_duh} we have
\begin{equation*}
\widehat{f}^1\left(\xi\right)=i\xi\sumd_{l=-\infty}^{\infty} \sumd_{d_1=-2}^{2}
\intd_{0}^{1}\left\{e^{it\xi M^*}\right\}_{k+d,k+l}
\delta M_{k+l,k+l+d_1}\left\{e^{i\left(1-t\right)\xi M^*}\right\}_{k+l+d_1,k-1} dt,
\end{equation*}
and by the spectral theorem and \eqref{e:M0_sp}
we have
\begin{equation}\label{e:f^1_m}
\widehat{f}^1\left(\xi\right)
=
\sumd_{l=-\infty}^{\infty} \sumd_{d_1=-2}^{2}
\delta M_{k+l,k+l+d_1}
\dfrac{e^{i\xi a}}{4c\pi^2} \intd_{-\pi}^{\pi}\intd_{-\pi}^{\pi}
\dfrac{e^{i\xi c\cos 2\phi}-e^{i\xi c\cos 2\psi}}{\cos 2\phi-\cos 2\psi}
e^{i\left(\left(l-d\right)\phi+\left(l+d_1+1\right)\psi\right)}d\phi d\psi
.
\end{equation}
Hence, integrating \eqref{e:f^1_m} with $\widehat{V'}$ we obtain the second term in \eqref{e:str_eq.m}
in the form
\begin{equation}\label{e:Lin_V`}
\dfrac{1}{2\pi}\intd_{-\infty}^{\infty}
\widehat{V'} \left(\xi\right) \widehat{f}^1\left(\xi\right) d\xi
=
\sumd_{l=-\infty}^{\infty} \sumd_{d_1=-2}^{2}
\delta M_{k+l,k+l+d_1}
\mathbf{B}_{l-d,l+d_1+1},
\end{equation}
where
\begin{equation}\label{e:B_ab_def}
\mathbf{B}_{\alpha,\beta}=\dfrac{1}{4\pi^2 c} \intd_{-\pi}^{\pi}du \intd_{-\pi}^{\pi}dv
\dfrac{V'\left(a+c\cos 2u\right)-V'\left(a+c\cos 2v\right)}{\cos 2u-\cos 2v} e^{i\left(\alpha u+ \beta v\right)}.
\end{equation}
Below we need some properties of these coefficients.
\begin{proposition}\label{p:B_ab}
For any function $V$ satisfying condition $C3$ coefficient $\mathbf{B}_{\alpha,\beta}$ depends only on
$\left|\alpha\right|+\left|\beta\right|$, i.e.
$\mathbf{B}_{\alpha,\beta} = \mathbf{B}_{\left|\alpha\right|+\left|\beta\right|}$, equals to zero when $\alpha$ or $\beta$
is odd and for any positive $\gamma$
\begin{equation}\label{e:B_ab}
\mathbf{B}_{2\gamma} =
\dfrac{1}{c} \sumd_{j-\gamma \in 2\mathbb{N}+1} v_{2j} = \underline{O}\left(\gamma^{-5/2}\right).
\end{equation}
\end{proposition}
Since $\delta M_{k,l}$ are bounded, the proposition yields that
the series in the r.h.s of \eqref{e:Lin_V`} is convergent.
Using \eqref{e:str_eq.m}, \eqref{e:null_p}, and \eqref{e:Lin_V`},
we obtain the equation valid for any $k$
\begin{multline}\label{e:Str_eq_s.1}
\dfrac{\widetilde{L}_{k,k+1}}{c}+
\left(\widetilde{L}_{k,k+1}+\widetilde{L}_{k,k-1}\right)
v_0
+
\sumd_{d,d_1=-2}^{2}\sumd_{l=-\infty}^{\infty} \widetilde{L}_{k,k+d} \mathbf{B}_{l-d,l+d_1+l} \delta M_{k+l,k+l+d_1}
+
r^{\left(n\right)}_k
\\=
\left(-1\right)^{k}\dfrac{\widetilde{\alpha}_{k}}{\widetilde{\rho}_{k}}
\left(
1+\dfrac{k}{n}
\right),
\end{multline}
for $$r^{\left(n\right)}_k=\dfrac{1}{2\pi} \intd_{-\infty}^{\infty}\widehat{V'}\left(\xi\right)\sumd_{d=-2}^{2}\widehat{r}\left(\xi;d\right)d\xi.$$
Set $$S_{k} = \sumd_{j=-10}^{10} \widetilde{x}_{k+j}^2.$$ Then $\widetilde{L}_{k,j}$ and $\delta M_{k,j}$
can be written up to the errors $\underline{O}\left(S_k\right)$ as
\begin{equation}\label{e:LM_kj}
\begin{array}{ll}
\delta M_{k,k-2}=
-\dfrac{b}{2}
\left(
\widetilde{x}_{k-1}
+
\widetilde{x}_{k}
\right)
,&
\widetilde{L}_{k,k-2}=
\left(-1\right)^{n+k}
\left(
\dfrac{c}{2}
+\dfrac{b}{2}\left(
\widetilde{x}_{k-1}
+
\widetilde{x}_{k}
\right)
\right)
\\
\delta M_{k,k-1}=
(-1)^{n+k-1} s \dfrac{c}{2}
\left(
\widetilde{x}_{k-1}
-
\widetilde{x}_{k+1}
\right)
,&
\widetilde{L}_{k,k-1}=
-sb-sa\widetilde{x}_{k}
+s\dfrac{c}{2}\left(
\widetilde{x}_{k-1}
+
\widetilde{x}_{k+1}
\right)
\\
\delta M_{k,k}=
b
\left(
\widetilde{x}_{k}
+
\widetilde{x}_{k+1}
\right)
,&
\widetilde{L}_{k,k}=0
\\
\delta M_{k,k+1}=
(-1)^{n+k} s \dfrac{c}{2}
\left(
\widetilde{x}_{k}
-
\widetilde{x}_{k+2}
\right)
,&
\widetilde{L}_{k,k+1}=
sb+sa\widetilde{x}_{k+1}
-s\dfrac{c}{2}\left(
\widetilde{x}_{k}
+
\widetilde{x}_{k+2}
\right)
\\
\delta M_{k,k+2}=
-\dfrac{b}{2}
\left(
\widetilde{x}_{k+1}
+
\widetilde{x}_{k+2}
\right)
,&
\widetilde{L}_{k,k+2}=
\left(-1\right)^{n+k+1}
\left(
\dfrac{c}{2}
+\dfrac{b}{2}\left(
\widetilde{x}_{k+1}
+
\widetilde{x}_{k+2}
\right)
\right)
\end{array}
\end{equation}
Replacing $\widetilde{L}_{i,j}$ and $\delta M_{i,j}$ by
their first order approximations and reminders we get a system of equations with respect to $\widetilde{x}_j$.
\begin{lemma}\label{l:r_k_est}
Under conditions C1-C3
equations \eqref{e:Str_eq_s.1} for $\left\{\widetilde{x}_{k}\right\}_{\left|k\right| \leq \varepsilon_1 n}$
have the form
\begin{equation}\label{e:Str_eq_A}
\sumd_{l}A_{k-l}\widetilde{x}_{l} = \dfrac{b}{c} \dfrac{k}{n} +
R_{k},
\end{equation}
and the Toplitz matrix $A$ has the symbol
\begin{equation}\label{e:Str_eq.sym}
\delta\left(\phi\right)=\sumd A_l e^{il\phi}=\dfrac{\sqrt{1-c\cos^2\left(\phi/2\right)}}{\pi\sqrt2} P \left(-2\arcsin\left(\sin\dfrac{\theta}{2}\cos\dfrac{\phi}{2}\right)\right).
\end{equation}
Reminder $R_k$ satisfies the bound
\begin{equation}\label{e:R_k_est_1}
R_k \leq C \left(\dfrac{k^2}{n^2}+S_k+\sumd S_{k+2l} \left|l\right|^{-5/2}\right)+r_k^{\left(n\right)},
\end{equation}
where
\begin{equation}\label{e:r_k_est}
\begin{array}{c}
\left|r^{\left(n\right)}_k\right|^2
\leq C \intd_{0}^{\infty} e^{-y} y^5 \left\| r_{k,y} \right\|_2^2 dy,
\\
r_{k,y} \left(x\right) = \min\left\{y^{-1}\sumd_{s=-\infty}^{\infty}
\left|
R^{\star}_{2s,0}
\right|^2
S_{k+2s},
\quad
\mathrm{dist}^{-1} \left(z,\left[-1,1\right]\right)
+
y^{-1}\Im R^{\star}_{0,0}
\right\},
\end{array}
\end{equation}
where $\left\|\cdot \right\|_2$ means the $L_2\left(\mathbb{R}\right)$ norm,
and $R^{\star}\left(z\right) = \left(M^{\star}-zI\right)^{-1},\quad z=-x+iy.$
\end{lemma}

Equations \eqref{e:Str_eq_A} are true only for $\left|k\right| \leq \varepsilon_1 n$.
But we  extend them to all $k \in \mathbb{Z}$
redefining  \eqref{e:Str_eq_A} by the identities.
All bounds for the reminders will be valid because the reminder for $k \sim n$ is of order  $1$.

Now we  start from the bound \eqref{e:NullApp} and using  the perturbation theory method
obtain \eqref{e:MainAs} in a few steps. Suppose  we have proved the bound
\begin{equation}\label{e:x_as}
\widetilde{x}_{m} =
\underline{O} \left( \log^{\beta}n\left(n^{-\alpha} +
\left|
m/n\right|^{\gamma}\right)\right).
\end{equation}
Note that on the first step we have it with $\alpha=1/4,\, \beta=1/2,\,\gamma=1/2$. This bound yields that
\begin{equation}\label{e:S_bound}
S_{k+2s} \leq \log^{2\beta} n \left(n^{-2\alpha}+n^{-2\gamma}+\min\left\{\left|
k/n\right|^{2\gamma}
+\left|
s/n\right|^{2\gamma},\,1\right\}\right).
\end{equation}
To estimate the reminder $R_k$ we must  estimate  $r_{k,y}\left(x\right)$ first. For this aim we
need some properties of $R^{\star}$. From the spectral theorem we have
\begin{equation}\label{e:R_jl}
R^{\star}_{j,l}=G_{j-l}:=\dfrac{1}{2\pi}\intd_{0}^{2\pi}
\dfrac{e^{i\left(j-l\right)\phi} d\phi}{a-z+c\cos 2\phi}.
\end{equation}
Using this representation we will estimate the sums in the r.h.s. of \eqref{e:r_k_est}.
\begin{proposition}\label{p:G_s s^p}
Let $G_l$ be defined in \eqref{e:R_jl}. Then for any integer $p$
\begin{equation}\label{e:G_s_est}
\intd_{-\infty}^{\infty}
\left(
\sumd_{s=0}^{\infty}
\left|
G_{2s} \left(-x+iy\right)
\right|^2
s^p
\right)^2
dx
\leq
C_p \dfrac{\max\left\{1,\log 1/y\right\}}{y^{2p+2}}.
\end{equation}
\end{proposition}
To estimate $r_k^{\left(n\right)}$ we use the second bound from the r.h.s of \eqref{e:r_k_est}
for small $y$ and the first bound for sufficiently big $y$. Combining \eqref{e:r_k_est} with \eqref{e:S_bound}  we get
\begin{multline*}
r_{k,y} \left(x\right)\leq \dfrac{C_{\alpha,\beta,\gamma}\log^{2\beta} n}{y}
\cdot
\\
\left\{
\begin{array}{ll}
\sumd_{s=0}^{\infty}
\left|
G_{2s} \left(-x+iy\right)
\right|^2
\left(
n^{-2\alpha}+n^{-2\gamma} +
\min\left\{\left|
k/n\right|^{2\gamma},\,1\right\}
+\left|
s/n\right|^{2\gamma}
\right) ,& y\geq n^{-\delta}
\\
\Im G_0,& y\leq n^{-\delta},
\end{array}
\right.
\end{multline*}
where $\delta$ we will choose on every step. Integrating the square of the last equation over $x$
and using \eqref{e:G_0_est} and  \eqref{e:G_s_est}, we obtain
\begin{equation*}
\left\|r_{k,y} \left(x\right)\right\|^2_2 \leq \dfrac{C'_{\alpha,\beta,\gamma}\log^{4\beta+1} n}{y^4}
\cdot
\left\{
\begin{array}{ll}
\left(
n^{-4\alpha}+n^{-4\gamma} +
\min\left\{\left|
k/n\right|^{4\gamma},\,1\right\}
+\left(ny\right)^{-4\gamma}
\right) ,& y\geq n^{-\delta}
\\
y^2\log 1/y,& y\leq n^{-\delta}
\end{array}
\right.
\end{equation*}
The last bound and \eqref{e:r_k_est} imply
\begin{equation}\label{e:r_k_n}
\left|r^{\left(n\right)}_k\right|^2 \leq C_{\alpha,\beta,\gamma,\delta}
\log^{4\beta+2} n
\left(
n^{-4\delta}+n^{-4\alpha}+n^{-4\gamma} +
\min\left\{\left|
k/n\right|^{4\gamma},\,1\right\}
+
n^{-4\gamma+\delta\left(4\gamma-2\right)}
\right).
\end{equation}
Now we estimate $R_k$. The term $\sumd_l S_{k+2l}l^{-5/2}$ can be estimate using \eqref{e:S_bound}.
\begin{eqnarray*}
\sumd_l S_{k+2l}l^{-5/2}& = &\sumd_{l \leq n}+\sumd_{l \geq n} \\
&\leq&
C'_{\alpha,\beta,\gamma,\delta}
\log^{2\beta} n
\left(
n^{-2\alpha}+n^{-2\gamma} +
\min\left\{\left|
k/n\right|^{2\gamma},\,1\right\}
+n^{-3/2}
\right).
\end{eqnarray*}
Combining the above relations finally we can estimate the reminder $R_k$:
$$
R_k = \underline{O} \left(
k^2/n^2+ n^{-3/2}
+ \log^{2\beta+1}n\left(n^{-2\delta}+ n^{-2\alpha} +
\min\left\{\left|
k/n\right|^{2\gamma},\,1\right\} +
n^{-2\gamma+\delta\left(2\gamma-1\right)}
 \right)\right).
$$
To find
$\widetilde{x}_j$ we find the inverse matrix of $\left\{ A_{k-l} \right\}_{k,l=-\infty}^{\infty}$.
By the spectral theorem
$$
\left(A^{-1}\right)_{k,l}=\left(A^{-1}\right)_{k-l} = \dfrac{1}{2\pi}\intd_{-\pi}^{\pi}
\dfrac{e^{i(k-l)\phi}}{\delta\left(\phi\right)} d\phi.
$$
From the conditions C1-C2 $P\left(\lambda\right)>C>0$ on the interval $[-\theta,\theta]$,
therefore the solution of the \eqref{e:Str_eq_A} can be found as
\begin{equation}\label{e:Str_eq_s.6}
\widetilde{x}_{k}=\dfrac{b}{c}\sumd_{m} \left( \dfrac{m+k}{n} + R_{m+k}\right)
\left(A^{-1}\right)_m
=
\dfrac{\pi \sqrt2 }{P\left(\theta\right) \sin\theta/2} \dfrac{k}{n}+
\sumd_{m} R_{m+k}\left(A^{-1}\right)_m
\end{equation}
Since $P\left(\lambda\right)$ is a twice differentiable function we have
$\sumd_{m}\left|A_{m}^{-1}\right|^2 m^4 \leq C.$ Therefore
\begin{multline*}
\left| \sumd_{m} R_{m+k} A^{-1}_m \right| / \log^{2\beta+1} n
\\
= \underline{O}\left( \sumd_{m} \left|A^{-1}_m\right|
\left(
n^{-2\delta} + n^{-2\alpha}+ n^{-3/2}+
n^{-2\gamma+\delta\left(2\gamma-1\right)}
+\min \left\{\dfrac{k^2}{n^2},1\right\}
+\min \left\{\left|\dfrac{k}{n}\right|^{2\gamma},1\right\}
\right)\right)
\\
+
\underline{O}
\left(\sumd_{|m|\leq n}
\left|A^{-1}_m\right| \left(\left|\dfrac{m}{n}\right|^2+\left|\dfrac{m}{n}\right|^{2\gamma}\right)
\right)
+
\underline{O}
\left(\sumd_{|m|\geq n}
\left|A^{-1}_m\right|\right)
\\
=
\underline{O}\left(
n^{-2\delta} + n^{-2\alpha}+ n^{-3/2}+
n^{-2\gamma+\delta\left(2\gamma-1\right)}
+\min \left\{\dfrac{k^2}{n^2},1\right\}
+\min \left\{\left|\dfrac{k}{n}\right|^{2\gamma},1\right\}
\right),
\end{multline*}
where we use
$$
\sumd_{|m|\leq n}
\left|A^{-1}_m\right| \left|\dfrac{m}{n}\right|^{2\gamma}
=
\sumd_{|m|\leq n}
\left|A^{-1}_m\right| m^2 \cdot \left|\dfrac{m}{n}\right|^{2\gamma}m^{-2} \leq Cn^{-3/2}.
$$
Using the last inequality for $\widetilde{x}$
with $\alpha=1/4,\, \beta=1/2,\, \gamma=1/2,\, \delta=1/4$ we obtain the bound \eqref{e:x_as} with
$\alpha=1/2,\, \beta=2,\, \gamma=1$. Taking $\delta=1/2$ we have $\alpha=1,\, \beta=5,\, \gamma=2$ and finally
with $\delta=2/3$ we get $\alpha=4/3,\, \beta=11,\, \gamma=2$.
\begin{equation*}
\widetilde{x}_k =\dfrac{\pi \sqrt2 }{P\left(\theta\right) \sin\theta/2} \dfrac{k}{n}+
\underline{O} \left(\log^{11} n \left(n^{-4/3}+
\dfrac{k^2}{n^2}\right)\right).
\end{equation*}

\section{Auxiliary results.}\label{s:Auxillary}
\textit{Proof of Proposition~\ref{p:Rho}.} To prove
\eqref{e:Rho_a} we denote
\begin{equation*}
D_{\theta}=\left\{z : \left|z\right|=1 , \mathrm{arg} z \in [-\theta,\theta]
\right\}.
\end{equation*}
Consider two functions defined on $D_{\theta}$
\begin{equation}\label{d:FG}
F\left(e^{i \lambda}\right)=\dfrac{i-\left(V\left(\cos \lambda\right)\right)'}{2 \pi
i}, \quad G\left(e^{i\lambda}\right)=\rho\left(\lambda\right).
\end{equation}
From \eqref{e:V_Rho} we obtain
\begin{equation*}\label{e:Arc_Eq}
F\left(\xi\right)=\dfrac{1}{\pi i} v.p.\intd_{\mathfrak{L}}
\dfrac{G\left(\zeta\right)}{\zeta-\xi}d\zeta,
\end{equation*}
where $\mathfrak{L}$ is $D_{\theta}$ with the direction from
$\zeta_0=e^{-i\theta}$ to $\overline{\zeta_0}$.
Then by the standard method (see [\onlinecite{Mu:53}]) we get
\begin{equation}\label{e:Sol_Phi}
G\left(z\right)=\dfrac{X\left(z\right)}{2\pi i}
v.p.\intd_{\mathfrak{L}}\dfrac{F\left(\zeta\right)}{\zeta -z}
\dfrac{d\zeta}{{X^{+}\left(\zeta\right)}}.
\end{equation}
where
\begin{equation}\label{e:Sol_H}
X\left(z\right)=\sqrt{\left(z-\zeta_0\right)\left(z-\overline{\zeta_0}\right)}.
\end{equation}
It is easy to see that $X^{+}\left(e^{i \lambda}\right)=\sqrt{2}
e^{i \lambda/2} \chi \left(\lambda\right)$. Using the identity \eqref{e:Int_sq}
we obtain that $v.p. \intd_{\mathfrak{L}}
\dfrac{d\zeta}{\left(\zeta-\zeta_1\right)X^{+}\left(\zeta\right)}=0$.
From the last relation and \eqref{d:FG} we obtain \eqref{e:Rho_a}.
\begin{flushright}
$\blacksquare$
\end{flushright}
\textit{Proof of Lemma~\ref{l:V_R}.} Consider the function
\begin{equation}\label{d:F}
F\left(\lambda\right)=\dfrac{1}{4\pi^2} \intd_{-\theta}^{\theta}
\cot \dfrac{\lambda-\mu}{2} \dfrac{P\left(\mu\right)(\cos \mu - \cos
\theta)-P\left(\lambda\right)(\cos \lambda - \cos \theta)\dfrac{\cos
\mu/2}{\cos \lambda/2}}{\sqrt{\cos \mu - \cos \theta}} d\mu.
\end{equation}
We prove first that $F$ coincides with $V'$ on $\sigma$. For
this aim we compute the integral
\begin{multline*}
I\left(\lambda\right)=%
v.p.\intd_{-\theta}^{\theta} \cot \dfrac{\lambda-\mu}{2}\dfrac{\cos
(\mu/2) d\mu}{\chi\left(\mu\right)}=%
v.p.\intd_{-\theta}^{\theta} \dfrac{\sin \lambda}{\cos \mu
- \cos \lambda}\dfrac{\cos
(\mu/2 )\, d\mu}{\chi\left(\mu\right)}
\\
=\dfrac{\sin \lambda}{\sqrt{2}}
v.p.\intd_{-\sin(\theta/2)}^{\sin(\theta/2)} \dfrac{dt}{\left(\sin^2
\lambda/2-t^2\right)\sqrt{\sin^2 \theta/2-t^2}}
=2\pi\dfrac{\cos \left(\lambda/2\right) \hbox{sign} \lambda}{\sqrt{\cos
\theta-\cos \lambda}}\mathbf{1}_{\left[-\theta,\theta\right]^c},
\end{multline*}
where we have used \eqref{e:Int_sq}.
This relation and \eqref{e:V_Rho} imply that
$F\left(\lambda\right)=\left(V\left(\cos \lambda\right)\right)'$ for $\lambda \in
\sigma$. Since $P$ is a twice differentiable function, we obtain
that $F$ is a differentiable function. Then
\begin{equation*}
\left.F\left(\lambda\right)-V'\left(\lambda\right)\right|_{\lambda=\theta-0}=0
\Rightarrow
\left.F'\left(\lambda\right)-V''\left(\lambda\right)\right|_{\lambda=\theta-0}=0
\Rightarrow F'\left(\theta\right)=V''\left(\theta\right)
\end{equation*}
The function
$F-V'$ is a differentiable on  $\sigma_{\varepsilon}$ and equals $0$
with its first derivative on  $\sigma$. Hence we have
$F\left(\lambda\right)-\left(V \left(\cos \lambda\right)\right)'=O\left(\left|\cos
\lambda -\cos \theta\right|\right)$ for $\lambda \in \sigma_{\varepsilon}\setminus \sigma$.
\begin{flushright}
$\blacksquare$
\end{flushright}
\textit{Proof of Proposition~\ref{p:Str_eq}.}
From the identity
$$
\intd_{-\pi}^{\pi}
\left(\chi_{k}^{\left(n\right)} \left(\lambda\right)
\overline{\chi_{k-1}^{\left(n\right)}\left(\lambda\right)}
e^{-nV\left(\cos\lambda\right)}\right)' d\lambda
=0,
$$
and orthogonality
$\left \langle
\frac{d}{d\lambda} \chi_{k-1}^{\left(n\right)} \left(\lambda\right)  , \,
\chi_{k}^{\left(n\right)}\left(\lambda\right)
\right \rangle=0$ we get
\begin{equation}\label{e:Str_eq_pr.1}
\intd_0^{2\pi} \sin \lambda V' \left(\cos \lambda\right)
\chi_{k}^{\left(n\right)} \left(\lambda\right)
\overline{\chi_{k-1}^{\left(n\right)}\left(\lambda\right)}
e^{-nV\left(\cos\lambda\right)} d\lambda
= -\dfrac{1}{n} \left \langle
\dfrac{d }{d\lambda}\chi_{k}^{\left(n\right)},
\chi_{k-1}^{\left(n\right)}
\right \rangle.
\end{equation}
To compute the r.h.s of \eqref{e:Str_eq_pr.1}
we use the definition of orthogonal polynomials \eqref{d:Chi_k} and some properties of polynomials $P_{k}^{\left(n\right)}$ and $Q_{k}^{\left(n\right)}$. We assume that $k$ is odd. For even $k$ the proof is similar. From the definition of $\chi_{k}^{\left(n\right)}$ we obtain
\begin{multline*}
I_{k,n}:=
\left \langle
\dfrac{d \chi_{2k+1}^{\left(n\right)} \left(\lambda\right) }{d\lambda} , \,
\chi_{2k}^{\left(n\right)}\left(\lambda\right)
 \right \rangle =
\left \langle
\dfrac{d}{d\lambda}
\left(e^{-ik\lambda} P_{2k+1}^{\left(n\right)} \left(\lambda\right)\right), \,
e^{-ik\lambda} Q_{2k}^{\left(n\right)} \left(\lambda\right)
 \right \rangle=
 \\
 =
-ik \left \langle
e^{-ik\lambda} P_{2k+1}^{\left(n\right)} \left(\lambda\right)
,\,
e^{-ik\lambda} Q_{2k}^{\left(n\right)} \left(\lambda\right)
\right \rangle
+
\left \langle
\dfrac{d}{d\lambda}
P_{2k+1}^{\left(n\right)} \left(\lambda\right)
,\,
Q_{2k}^{\left(n\right)} \left(\lambda\right)
\right \rangle.
\end{multline*}
From the definition of $P_{k}^{\left(n\right)}$ we have
\begin{equation}\label{e:PQ_ort}
P_k^{\left(n\right)}\left(\lambda\right) \perp e^{im\lambda},
m=0,\ldots,k-1; \quad Q_k^{\left(n\right)}\left(\lambda\right) \perp
e^{im\lambda}, m=1,\ldots,k.
\end{equation}
Therefore the first term in the above relation is zero. Hence
\begin{equation*}
I_{k,n}=
i \left(2k+1\right) c_{2k+1,2k+1}^{\left(n\right)}
\left \langle
e^{i\left(2k+1\right)\lambda},\,
Q_{2k}^{\left(n\right)} \left(\lambda\right)
\right \rangle.
\end{equation*}
Now using \eqref{e:PQ_ort} one more time and equality $\left\|P_{k}^{\left(n\right)} \right\|=1$ we obtain
\begin{multline}\label{e:Str_odd}
I_{k,n}=
i \left(2k+1\right)
\left \langle
P_{2k+1}^{\left(n\right)} \left(\lambda\right) -
c_{2k+1,0}^{\left(n\right)} ,\,
Q_{2k}^{\left(n\right)} \left(\lambda\right)
\right \rangle
=
\\
=- i \left(2k+1\right) c_{2k+1,0}^{\left(n\right)}
\left \langle
1,\,
Q_{2k}^{\left(n\right)} \left(\lambda\right)
\right \rangle
=
- i \left(2k+1\right) c_{2k+1,0}^{\left(n\right)}
\left \langle
P_{2k}^{\left(n\right)} \left(\lambda\right),\,
e^{i2k\lambda}
\right \rangle
=
\\
=
-i\left(2k+1\right)\dfrac{c_{2k+1,0}^{\left(n\right)}}{c_{2k,2k}^{\left(n\right)}}
=
-i\left(2k+1\right)\dfrac{\alpha_{2k+1}^{\left(n\right)}}{\rho_{2k+1}^{\left(n\right)}}
\end{multline}
Relations \eqref{e:Str_eq_pr.1} and \eqref{e:Str_odd} give us \eqref{e:Str_eq}.
\begin{flushright}
$\blacksquare$
\end{flushright}

\textit{Proof of Proposition~\ref{p:v_1_v_0}.}
We prove the proposition, using  \eqref{e:V_Rho}. Consider the difference $v_2-2v_0$
and its integral representation.
\begin{equation*}
v_2-2v_0=\dfrac{1}{\pi}\intd_{-\pi}^{\pi} V'\left(a+c\cos2\phi\right) \left(\cos2\phi-1\right) d\phi.
\end{equation*}
Changing the variables with $\sin \dfrac{\lambda}{2}= \sin \dfrac{\theta}{2}\sin \phi$,
we get
\begin{equation*}
v_2-2v_0 = \dfrac{1}{\pi c^{3/2}} \intd_{\sigma} -V'\left(\cos \lambda\right) \sin\lambda \dfrac{\sin \lambda/2}{\cos \phi} d\lambda.
\end{equation*}
And finally, using  \eqref{e:V_Rho}, we obtain
\begin{equation*}
v_2-2v_0 = \dfrac{1}{c\pi} \intd_{\sigma} d\mu \rho \left(\mu\right) v.p. \intd_{\sigma}  \cot \dfrac{\lambda-\mu}{2} \dfrac{\sin \lambda/2 d\lambda}{\sqrt{\sin^2 \theta/2 -\sin^2 \lambda/2}}
=\dfrac{2}{c} \intd_{\sigma}  \rho \left(\mu\right)d\mu = \dfrac{2}{c}.
\end{equation*}
~
\begin{flushright}
$\blacksquare$
\end{flushright}
\textit{Proof of Proposition \ref{p:B_ab}}
We  prove first \eqref{e:B_ab} for $V'\left(a+c\cos 2u\right) = \cos 2ku$ with $k\geq 0$.
$\mathbf{B}$ is symmetric over changing
$\alpha \rightarrow -\alpha,\,\beta \rightarrow -\beta,\,\alpha \leftrightarrow \beta$.
Hence, we can assume $\beta\geq\alpha\geq0$. For odd $\alpha$ or $\beta$ we will have $0$, because of
the ratio under the integral in the \eqref{e:B_ab_def} does not change after the shift $u \to \pi+u$ and $v \to \pi+v$.
Therefore we change $\alpha\to 2\alpha$ and $\beta\to 2\beta$. And finally, changing $2u\to u$ and
$2v\to v$ in  \eqref{e:B_ab_def}, we obtain
\begin{equation*}
\mathbf{B}_{2\alpha,2\beta}^{\left(k\right)}:=\dfrac{1}{4\pi^2 c} \intd_{-\pi}^{\pi}du \intd_{-\pi}^{\pi}dv
\dfrac{\cos ku-\cos kv}{\cos u-\cos v} \cos \alpha u \cos \beta v.
\end{equation*}
To compute this integral we use two evident relations
\begin{equation}\label{e:Int_sin}
\intd_{-\pi}^{\pi} \dfrac{\sin l u}{\sin u} du= 2\pi \mathrm{sign}\, l \cdot
\left\{
\begin{array}{cc}
1, & \mathrm{l \quad is \quad odd},
\\
0, & \mathrm{l \quad is \quad even}.
\end{array}
\right.
\end{equation}
\begin{equation}\label{e:Int_cos}
\intd_{-\pi}^{\pi} \dfrac{\cos l u - \cos l v}{\cos u-\cos v} dv= 2\pi \dfrac{\sin l u}{\sin u}.
\end{equation}
Now to apply \eqref{e:Int_sin} and \eqref{e:Int_cos} we transform integral for $\mathbf{B}_{2\alpha,2\beta}^{\left(k\right)}$ as follows
\begin{eqnarray*}
\mathbf{B}_{2\alpha,2\beta}^{\left(k\right)}
&=&
\dfrac{1}{4\pi^2 c} \intd_{-\pi}^{\pi}du \intd_{-\pi}^{\pi}dv
\left( \cos ku-\cos kv \right)
\dfrac{\cos \alpha u - \cos \alpha v}{\cos u-\cos v} \cos \beta v
\\
&+&
\dfrac{1}{4\pi^2 c} \intd_{-\pi}^{\pi}du \intd_{-\pi}^{\pi}dv
\dfrac{\cos ku-\cos kv}{\cos u-\cos v} \cos \alpha v \cos \beta v
\end{eqnarray*}
We write the first integral as a sum of two parts: one with $\cos ku$ and
the other with $\cos kv$. Then in the second part we integrate over $u$ using \eqref{e:Int_cos}.
In the second integral  above  we similarly integrate over $u$ and use \eqref{e:Int_cos}.
Then we obtain
\begin{eqnarray*}
\mathbf{B}_{2\alpha,2\beta}^{\left(k\right)}
&=&
\dfrac{1}{4\pi^2 c} \intd_{-\pi}^{\pi}du \cos ku \intd_{-\pi}^{\pi}dv
\dfrac{\cos \alpha u - \cos \alpha v}{\cos u-\cos v} \cos \beta v
\\
&+&
\dfrac{1}{2\pi c} \intd_{-\pi}^{\pi}dv
\dfrac{\sin kv-\sin \alpha v}{\sin v} \cos \alpha v \cos \beta v.
\end{eqnarray*}
Note that the function $f_u\left(v\right):=\dfrac{\cos \alpha u - \cos \alpha v}{\cos u-\cos v}$ can
be represented as a trigonometric polynomial of $v$ with degree not bigger than $\alpha-1$,
therefore the first integral in the above relation is equal to $0$ if $\beta \geq \alpha$.
And finally we have

\begin{eqnarray}
\mathbf{B}_{2\alpha,2\beta}^{\left(k\right)}
&=&
\dfrac{1}{4\pi c} \intd_{-\pi}^{\pi}
\dfrac{ \sin \left( k-\alpha-\beta \right) v + \sin \left( k-\alpha+\beta \right) v}{\sin v} \, dv
\notag
\\
&=&
\left\{
\begin{array}{cc}
c^{-1}, & k - \alpha -\beta \in 2\mathbb{N}+1
\\
0, & otherwise
\end{array}
\right.
.
\label{e:B_ab_pr}
\end{eqnarray}
Now using \eqref{e:B_ab_pr} and the Fourier expansion $V'\left(a+c\cos2\phi\right)=
\sumd_{k=0}^{\infty} v_{2k} \cos 2k \phi$, we obtain the first equality in \eqref{e:B_ab}.
To estimate $\mathbf{B}_{2\alpha,2\beta}$ we use the Schwarz inequality
\begin{equation*}
\left|\sumd_{k-\alpha-\beta \in 2\mathbb{N}+1} v_{2k} \right| \leq \sqrt{\sumd_{k>2\alpha+2\beta} v^2_k k^{6}}
\sqrt{\sumd_{k > 2\alpha+2\beta} \dfrac{1}{k^{6}}} =\underline{O}\left(\left(\alpha+\beta\right)^{-5/2}\right).
\end{equation*}
\begin{flushright}
$\blacksquare$
\end{flushright}
\textit{Proof of Lemma \ref{l:r_k_est}}
Using \eqref{e:LM_kj} in \eqref{e:Str_eq_s.1} we obtain
\begin{eqnarray}
\dfrac{b}{c} &+&\dfrac{a}{c}\widetilde{x}_{k+1}-\dfrac{1}{2}\left(\widetilde{x}_{k}+\widetilde{x}_{k+2}\right) +
v_0
\left(
\dfrac{c}{2}\widetilde{x}_{k-1} +\dfrac{1+a}{2}\left(-\widetilde{x}_{k}+\widetilde{x}_{k+1}\right)
-\dfrac{c}{2}\widetilde{x}_{k+2}
\right) \notag
\\
&+&
\sumd_{l=-\infty}^{\infty}
\bigg[
\dfrac{c^2}{4} \left(\mathbf{B}_{\left|4l+2\right|} - \mathbf{B}_{\left|4l-2\right|}\right)
\left(\widetilde{x}_{k+2l-1}-\widetilde{x}_{k+2l+1}\right) \notag
\\
&+&
\dfrac{c^2}{4} \left(\mathbf{B}_{\left|4l+4\right|} - \mathbf{B}_{\left|2l-2\right|+\left|2l+2\right|}\right)
\left(\widetilde{x}_{k+2l+2}-\widetilde{x}_{k+2l}\right) \notag
\\
&-&
\dfrac{b^2}{2} \left(\mathbf{B}_{\left|4l-2\right|} - \mathbf{B}_{\left|4l-4\right|}\right)
\left(\widetilde{x}_{k+2l-2}+\widetilde{x}_{k+2l-1}\right)\notag
\\
&-&
\dfrac{b^2}{2} \left(\mathbf{B}_{\left|4l+2\right|} - \mathbf{B}_{\left|2l-2\right|+\left|2l+2\right|}\right)
\left(\widetilde{x}_{k+2l}+\widetilde{x}_{k+2l+1}\right) \notag
\\
&+&
b^2\left(\mathbf{B}_{\left|4l\right|} - \mathbf{B}_{\left|4l-2\right|}\right)
\left(\widetilde{x}_{k+2l-1}+\widetilde{x}_{k+2l}\right)
\bigg]
=
\dfrac{b}{c} \left(1+\dfrac{k}{n}\right)
-
\dfrac{1}{c} \widetilde{x}_{k}
+R_k,\label{e:Str_eq_s.2}
\end{eqnarray}
where $R_k$ accumulates the errors of \eqref{e:Str_eq_s.1} and \eqref{e:LM_kj}. It is easy to see that
\begin{equation*}
\left|R_k\right| \leq C
\left(
\dfrac{k^2}{n^2}+ S_{k} +
\sumd_{l=-\infty}^{\infty}
\left(S_{k}+S_{k+2l}\right)
\left(
\left|\mathbf{B}_{\left|4l-4\right|}\right|
+
\ldots
+
\left|\mathbf{B}_{\left|4l+4\right|}\right|
\right)
\right)
+r^{\left(n\right)}_k.
\end{equation*}
Recollecting terms with $\widetilde{x}_j$ we obtain the system \eqref{e:Str_eq_s.2} in the form \eqref{e:Str_eq_A}.
Definitions \eqref{e:Str_eq_A}, \eqref{e:Str_eq.sym} and system \eqref{e:Str_eq_s.2} imply

\begin{multline}\label{e:Str_eq.sym2}
\delta\left(\phi\right) = A_0 + 2\sumd_{l=1}^{\infty}
A_l \cos\left(l\phi\right)
\\
=
- av_0
+\dfrac{1-\left(1+a\right)v_0}{2}
+cv_0\cos\phi
+
v_2\left(
\dfrac{c}{4}-\dfrac{c}{2}\cos 2\phi
+2a\cos\phi
+2a
\right)
\\
+
\sumd_{l=1} v_{4l}
\left(
\dfrac{c}{2} \left(\cos\left(2l+1\right)\phi - \cos\left(2l-1\right)\phi\right)
-2a\cos2l\phi -2a - 4a \left(\sumd_{k=1}^{2l-1}\cos k\phi\right)
\right)
\\
+
\sumd_{l=2} v_{4l-2}
\left(
\dfrac{c}{2} \left(\cos\left(2l-2\right)\phi - \cos\left(2l\right)\phi\right)
+2a\cos\left(2l-1\right)\phi +2a + 4a \left(\sumd_{k=1}^{2l-2}\cos k\phi\right)
\right)
\\
=
\left(c\sin\phi+2a\cot \dfrac{\phi}{2}\right)
\sumd_{l=1}^{\infty} \left(-1\right)^{l-1} \sin l\phi v_{2l}
-\dfrac{c}{4}v_2
+
v_0\left(
-a
-\dfrac{1+a}{2}
+c\cos\phi
\right)
+
\dfrac{1}{2}
\\
=
-2\left(a+c\sin^2\dfrac{\phi}{2}\right)
\left(
\sumd_{l=1}^{\infty} v_{2l} \left(-1\right)^l \dfrac{\sin l\phi}{\sin\phi} 2\cos^2\dfrac{\phi}{2}
+
v_0
\right).
\end{multline}
On the other hand, from the definition \eqref{d:Sqrt} we have
\begin{multline}\label{e:P_eq}
P\left(2\arcsin \left(\sin \dfrac{\theta}{2}\sin \dfrac{v}{2}\right)\right)
=
2\pi\sqrt{2\left(1-c\sin^2 \dfrac{v}{2}\right)}
\left(\left(1-\cos v\right)
\sumd_{l=1}^{\infty} v_{2l} \dfrac{\sin lv}{\sin v}
- v_0
\right).
\end{multline}
The equality can be obtained from \eqref{d:Sqrt} by the change of variables
$\sin \lambda/2 = \sin \theta/2 \sin u/2$ and
$\sin \mu/2 = \sin\theta/2 \sin v/2$
\begin{multline}\label{e:P_eq_pr}
P\left(\mu\right) = \intd_{-\theta}^{\theta}
\dfrac{V'\left(\lambda\right) - V'\left(\mu\right)}{\sin
\dfrac{\lambda-\mu}{2}} \dfrac{d\lambda}{\sqrt{\cos\lambda-\cos\theta}}
\\
=
\sqrt{2}
\intd_{-\pi}^{\pi}
\dfrac{
\sin v/2 \cos \mu/2V'\left(a+c\cos v\right)
-\sin u/2 \cos \lambda/2V'\left(a+c\cos u\right)
}
{
\sin u/2 \cos \mu/2
-
\sin v/2 \cos \lambda/2
}
\dfrac{du}{\cos \lambda/2}
\\
=
\sqrt{2}
\intd_{-\pi}^{\pi}
\dfrac{
\cos \mu/2
\left(
\sin^2 v/2 V'\left(a+c\cos v\right)
-\sin^2 u/2 V'\left(a+c\cos u\right)
\right)
}
{
\sin^2 u/2
-
\sin^2 v/2
}
du
\\
=
\sqrt{2}
\cos \mu/2
\left(
\left(1-\cos v\right)\intd_{-\pi}^{\pi}
\dfrac{V'\left(a+c\cos u\right)-V'\left(a+c\cos v\right)}{\cos u-\cos v} du
-2\pi v_0.
\right)
\end{multline}
Now, using the definition of $v_{2l}$ and equality
$$\intd_{-\pi}^{\pi}
\dfrac{\cos lu-\cos lv}{\cos u-\cos v} du = 2\pi \dfrac{\sin
lv}{\sin v},$$
we obtain \eqref{e:P_eq}.
Therefore,
\begin{equation}\label{e:sym_P}
\delta\left(\phi\right) = \dfrac{\sqrt{1-c\cos^2\dfrac{\phi}{2}}}{\pi\sqrt2}
P \left(-2\arcsin\left(\sin\dfrac{\theta}{2}\cos\dfrac{\phi}{2}\right)\right).
\end{equation}
To estimate the reminder we will use the inequality
\begin{equation}\label{e:Johanson}
\left|\intd_{-\infty}^{\infty} f\left(x\right) \overline{g\left(x\right)} dx\right|
\leq
\sqrt{
\dfrac{1}{2\pi\Gamma \left(2s\right)}
\intd_{0}^{\infty}e^{-y}y^{2s-1}\left\|f \star P_y\right\|^2_2 dy
}
\sqrt{
\intd_{-\infty}^{\infty}\left|\widehat{g}\left(\xi\right)\right|^2 \left(2\left|\xi\right|+1\right)^{2s} d \xi
},
\end{equation}
where $P_y\left(t\right)=\dfrac{y}{\pi\left(y^2 + t^2\right)}$, $g\left(x\right)$ any $s$~--times differentiable function with $g^{(s)} \in L_2$ and $f \in L_2$.
The inequality was proposed in [\onlinecite{Jo:98}]. It
 can be obtained from the  Schwartz inequality
\begin{multline*}
\left|\intd_{-\infty}^{\infty} f\left(x\right) \overline{g\left(x\right)} dx\right|=
\left|\dfrac{1}{2\pi}\intd_{-\infty}^{\infty} \widehat{f}\left(\xi\right) \overline{\widehat{g}\left(\xi\right)} d \xi\right| \\
\leq
\dfrac{1}{2\pi}
\left(\intd_{-\infty}^{\infty} \left|\widehat{f}\left(\xi\right)\right|^2 \left(2\left|\xi\right|+1\right)^{-2s}d\xi\right)^{1/2}
\left(\intd_{-\infty}^{\infty} \left|\widehat{g}\left(\xi\right)\right|^2 \left(2\left|\xi\right|+1\right)^{2s}d\xi\right)^{1/2},
\end{multline*}
and exact computation of the Fourier transform of $P_y$
\begin{equation*}
\widehat{P}_y\left(\xi\right)=\dfrac{1}{\pi}\intd_{-\infty}^{\infty} \dfrac{y}{y^2 + t^2} e^{-it\xi} dt = sign \left(y\right) e^{-\left|y\right|\left|\xi\right|}.
\end{equation*}
Now, using that
\begin{equation*}
\left\|f \star P_y\right\|^2_2 = \dfrac{1}{2\pi}\left\|\widehat{f} \widehat{P}_y \right\|^2_2,
\end{equation*}
and the definition of the $\Gamma$ function we obtain
\begin{equation*}
\dfrac{1}{\Gamma\left(2s\right)} \intd_{0}^{\infty} e^{-y} y^{2s-1}
\left|\widehat{P}_y\left(\xi\right) \right|^2 dy=
\dfrac{1}{\Gamma\left(2s\right)} \intd_{0}^{\infty} e^{-y\left(2\left|\xi\right|+1\right)} y^{2s-1} dy = \left(2\left|\xi\right|+1\right)^{-2s}.
\end{equation*}
Finally, combining the above relations, we get \eqref{e:Johanson}.
We use this inequality for $\widehat{V'}$ and $\widehat{r}\left(\xi;d\right)$
with $s=3$.
\begin{eqnarray*}
\left|
\dfrac{1}{2\pi}\intd \widehat{V'} \left(\xi\right)\widehat{r}\left(\xi;d\right)d\xi
\right|^2
&\leq&
C_{V} \intd_{0}^{\infty} e^{-y} y^5 \left\| r\left(d\right) \star P_y\right\|_2^2 dy
\\
&=&
C_{V} \intd_{0}^{\infty} e^{-y} y^5 \left\| \left(f-f^0-f^1\right) \star P_y\right\|_2^2 dy.
\end{eqnarray*}
On the other hand, for any self-adjoint linear operator $A$ and any linear $B$ we have
\begin{equation}\label{e:P_conv}
\begin{array}{c}
\dfrac{1}{2\pi} \intd e^{ix\xi} \widehat{P_y} \left(\xi\right) e^{iA\xi}   d\xi
= \dfrac{1}{\pi} \Im G_{A} \left(-x+iy\right), \quad
\\
\dfrac{1}{2\pi} \intd e^{ix\xi} \widehat{P_y} \left(\xi\right) e^{iA\xi t} B e^{iA\xi \left(1-t\right)}   d\xi
=
\dfrac{1}{\pi} \Im G_{A} \left(-x+iy\right)B G_{A} \left(-x+iy\right),
\end{array}
\end{equation}
where $G_{A} \left(z\right) = \left(A-z\right)^{-1}.$ Hence from \eqref{e:g_j_def} and \eqref{e:P_conv}
we have
\begin{equation}\label{e:f_P}
\left(f \star P_y \right)\left(x\right)=
\dfrac{1}{\pi}\Im
R_{k+d,k-1}\left(-x+iy\right),
\end{equation}
\begin{equation}\label{e:f^0_P}
\left(f^{0} \star P_y \right)\left(x\right)
=
\dfrac{1}{\pi}\Im
R^{\star}_{k+d,k-1}\left(-x+iy\right),
\end{equation}
\begin{equation}\label{e:f^1_P}
\left(f^{1} \star P_y \right)\left(x\right)=
\dfrac{1}{\pi}\Im
\left\{
R^{\star}\left(-x+iy\right)
\delta M
R^{\star}\left(-x+iy\right)
\right\}_{k+d,k-1}
.
\end{equation}
Using the resolvent identity,  \eqref{e:f_P}, \eqref{e:f^0_P}, and \eqref{e:f^1_P}, we obtain
\begin{multline}\label{e:r_P}
\left(r \star P_y \right)\left(x\right)=
\left(f \star P_y \right)\left(x\right)-
\left(f^0 \star P_y \right)\left(x\right)-
\left(f^1 \star P_y \right)\left(x\right)
\\
=
\Im
\left\{
R^{\star}\left(-x+iy\right)
\delta M
R\left(-x+iy\right)
\delta M
R^{\star}\left(-x+iy\right)
\right\}_{k+d,k-1}
.
\end{multline}
Now we are ready to estimate $r\left(d\right)\star P_y$.
\begin{multline}\label{e:r_P.est}
\left|
r\left(d\right) \star P_y
\right|
\leq
\left|
\langle
R^{\star}\left(-x+iy\right)
\delta M
R\left(-x+iy\right)
\delta M
R^{\star}\left(-x+iy\right)
e_{k-1},
e_{k+d}
\rangle
\right|
\\
\leq
\left\|R\right\|
\sqrt{
\left(
R^{\star}\left(-x+iy\right)
\delta M^2
R^{\star}\left(-x-iy\right)
\right)_{k+d,k+d}
}
\\
\cdot
\sqrt{
\left(
R^{\star}\left(-x-iy\right)
\delta M^2
R^{\star}\left(-x+iy\right)
\right)_{k-1,k-1}
}
.
\end{multline}
Both  roots can be estimated using the definition of $\delta M$
and \eqref{e:R_jl}.
Finally we obtain
\begin{equation}\label{e:r_P_est}
\left|
r\left(d\right) \star P_y
\right|
\leq
\left\|R\right\|
\sumd_{s=-\infty}^{\infty}
\left|
G_{2s}
\right|^2
S_{k+2s}
\end{equation}
For sufficiently small $y$ we will use another bound
\begin{multline}\label{e:r_P.est2}
\left|
r\left(d\right) \star P_y
\right|
\leq
\left|
\left(f \star P_y \right)\left(x\right)
\right|+
\left|
\left(f^0 \star P_y \right)\left(x\right)
\right|+
\left|
\left(f^1 \star P_y \right)\left(x\right)
\right|
\\
\leq
C
\left(
\left\|
R
\right\|
+
\left\|
R^{\star}
\right\|
+
\left\|
\delta M
\right\|
\left(
\left\{
R^{\star} \left(z\right)
R^{\star} \left(\overline{z}\right)
\right\}_{k+d,k+d}
+
\left\{
R^{\star} \left(z\right)
R^{\star} \left(\overline{z}\right)
\right\}_{k-1,k-1}
\right)
\right)
\\
\leq
C
\left(
\left\|
R
\right\|
+
\left\|
R^{\star}
\right\|
+
\sumd_{d=-2}^{d=2}
\left\{
R^{\star} \left(z\right)
R^{\star} \left(\overline{z}\right)
\right\}_{k+d,k+d}
\right)
.
\end{multline}
\begin{flushright}
$\blacksquare$
\end{flushright}
\textit{Proof of Proposition \ref{p:G_s s^p}}
It is easy to see that we can prove this proposition for even $p$. For odd $p$
it  follows from the Schwartz inequality. The definition \eqref{e:R_jl} implies
\begin{equation*}
\dfrac{1}{a-z+c\cos2\phi} = \sumd_{s} G_{2s}e^{-2is\phi}.
\end{equation*}
Hence
\begin{multline*}
\sumd_s \left|G_{2s}\right|^2 s^{2p}=
\sumd_s G_{2s} s^{2p} \dfrac{1}{2\pi}\intd_{-\pi}^{\pi}
\dfrac{e^{-2is\phi}}{a-\overline{z}+c\cos2\phi}d\phi
\\
=
\left(-\dfrac{1}{4}\right)^{p}
\dfrac{1}{2\pi}\intd_{-\pi}^{\pi}
\left(\dfrac{\partial^{2p}}{\partial \phi^{2p}} \dfrac{1}{a-z+c\cos2\phi}\right)
\dfrac{1}{a-\overline{z}+c\cos2\phi}d\phi
\\
=
C_p
\intd_{-\pi}^{\pi}
\left|\dfrac{\partial^{p}}{\partial \phi^{p}} \dfrac{1}{a-z+c\cos2\phi}\right|^2
d\phi
\leq
C'_p
y^{-2p}
\intd_{-\pi}^{\pi}
\dfrac{d\phi}{\left|a-z+c\cos2\phi\right|^2}
d\phi
\\
=
C''_p
y^{-2p-1}
\Im G_0
.
\end{multline*}
From the residue theorem we obtain
\begin{equation*}
G_{0}=\dfrac{2\zeta}{c\left(\zeta^2-1\right)}, \quad \mbox{where} \,
\zeta^2+2\dfrac{a-z}{c}\zeta+1=0\, \mbox{and} \, \left|\zeta\right| < 1.
\end{equation*}
Therefore,
\begin{equation*}
\left|G_{0}\right| = \dfrac{1}{\sqrt{\left|z-1\right|\left|z-\cos\theta\right|}}.
\end{equation*}
Finally, combining all above bounds, we obtain
\begin{multline}\label{e:G_0_est}
\intd_{-\infty}^{\infty}\left(\sumd_s \left|G_{2s}\right|^2 s^{2p}\right)^2 dx
\leq C'''_p y^{-4p-2} \intd_{-\infty}^{\infty}
\dfrac{dx}{\left|z-1\right|\left|z-\cos\theta\right|}
\\
\leq \widetilde{C}_p y^{-4p-2} \max\left\{\log 1/y,\, 1\right\}.
\end{multline}
\begin{flushright}
$\blacksquare$
\end{flushright}

\begin{acknowledgements}
 The author  is grateful to Prof. M.Shcherbina
 for  the problem statement and fruitful discussions.
 The author was partially supported by the Akhiezer fund.
\end{acknowledgements}

\end{document}